\documentclass[fleqn,usenatbib]{mnras}

\usepackage{newtxtext,newtxmath}

\usepackage[T1]{fontenc}

\DeclareRobustCommand{\VAN}[3]{#2}
\let\VANthebibliography\thebibliography
\def\thebibliography{\DeclareRobustCommand{\VAN}[3]{##3}\VANthebibliography}

\usepackage{graphicx}	
\usepackage{amsmath}	

\usepackage[dvipsname]{xcolor} 
\usepackage{array,multirow}  
\definecolor{auburn}{rgb}{0.43, 0.21, 0.1}

\usepackage{transparent}

\title[Short title, max. 45 characters]{MNRAS \LaTeXe\ template -- title goes here}
\title[Origin of the star-forming gas in \textsc{eagle} galaxies]{History of the gas fuelling  star formation in \textsc{eagle} galaxies}

\author{Laura Scholz-D\'{i}az}

\author[L. Scholz-D\'{i}az et al.]{
Laura Scholz-D\'{i}az$^{1,2}$, 
Jorge S\'anchez Almeida$^{1,2}$\thanks{E-mail: jos@iac.es}, 
\& Claudio Dalla Vecchia$^{1,2}$
\\
$^{1}$Instituto de Astrof\'\i sica de Canarias, La Laguna, Tenerife, E-38200, Spain\\
$^{2}$Departamento de Astrof\'\i sica, Universidad de La Laguna, Spain
}

\date{Accepted XXX. Received YYY; in original form ZZZ}

\pubyear{2021}

\begin{document}
\label{firstpage}
\pagerange{\pageref{firstpage}--\pageref{lastpage}}
\maketitle

\begin{abstract}
Theory predicts that cosmological gas accretion plays a fundamental role fuelling star formation in galaxies. However,  a detailed description of the accretion process to be used when interpreting observations is still lacking. Using the state-of-the-art cos\-mo\-log\-i\-cal hydrodynamical simulation \textsc{eagle}, we work out the chemical inhomogeneities arising in the disk of galaxies due to the randomness of the accretion process. In low-mass systems and outskirts of massive galaxies, low metallicity regions are associated with enhanced star-formation, a trend that reverses in the centers of massive galaxies. These predictions agree with the relation between surface density of star formation rate and metallicity observed in the local spiral galaxies from the  MaNGA survey. Then, we analyse the origin of the gas that produce stars at two key epochs, $z\simeq 0$ and $z\simeq 2$. The main contribution comes from gas already in the galaxy about 1\,Gyr before stars are formed, with a share from external gas that is larger at high redshift. The accreted gas may come from major and minor mergers, but also as  gravitationally unbound gas and from mergers with dark galaxies (i.e., haloes where more than 95\,\% of the baryon mass is in gas). We give the relative contribution of these sources of gas as a function of stellar mass ($8 \le \log[M_\star/{\rm M}_\odot] \le 11$). Even at $z=0$, some low-mass galaxies form a significant fraction of their total stellar mass during the last Gyr  from mergers with dark galaxies.
 %
\end{abstract}


\section{Introduction}
\label{intro}

Numerical simulations predict that cosmological metal-poor gas accretion drives the growth of disc galaxies  \citep[e.g.][]{2006MNRAS.368....2D,2009Natur.457..451D,2012RAA....12..917S,2012MNRAS.425..788G}. The gas that falls into galaxies and feeds star formation often corresponds to cold gas accretion \citep[e.g.][]{2005MNRAS.363....2K,2009Natur.457..451D,2011MNRAS.414.2458V}. This {\em cold mode} is expected to supply gas mainly through cold filamentary streams from the cosmic web  \citep[e.g.][]{2009Natur.457..451D,2012RAA....12..917S}, that do not get shock-heated when entering the galactic halo, and reach the galaxy disc directly ready to fuel star formation \citep[e.g.][]{2009ApJ...694..396B}. In this simple picture, this gas accretion mode dominates at early times and low halo masses, below $\sim 10^{12}\ \mathrm{M_{\odot}}$  \citep[e.g.][]{2003MNRAS.345..349B,2011MNRAS.414.2458V}. In contrast, the {\em hot mode} is dominant for higher mass haloes, in which the gas falling into the halo is shock-heated to approximately the halo virial temperature, requiring much longer timescales to cool, condense, and form stars \citep[e.g.][]{2003MNRAS.345..349B,2005MNRAS.363....2K}. Although galaxies also grow through mergers  \citep[e.g.][]{2011MNRAS.413..101G,2013seg..book....1K}, simulations indicate that gas accreted directly from the cosmic web dominates mergers when considering galaxies outside dense environments  \citep[e.g.][]{2011MNRAS.413.1373W,2012A&A...544A..68L,2013MSAIS..25...45C,2011MNRAS.414.2458V}.  Cold-mode gas is expected to have considerably lower metallicities than galaxy outflows driven by star formation feedback or AGN activity  \citep{2012MNRAS.423.2991V}.

  Despite the clear-cut theoretical prediction, observational evidence for cosmic gas accretion sustaining star formation remains indirect \cite[for a recent review, see][]{2017ASSL..430...67S}. This gas is elusive from an observational point of view given the highly complicated nature of the intergalactic medium (IGM), predicted to be tenuous and multiphase. To complicate interpretation further, in the circum-galactic medium (CGM) the incoming gas gets entangled with metal-rich recycled material ejected from the galaxy. Moreover,  pure gas accretion events and gas-rich minor mergers are difficult to distinguish both in numerical simulations and in observations \citep[e.g.,][]{2015ApJ...810L..15S}. 

  An example of indirect observational evidence for metal-poor gas accretion is given by the correlation between star formation rate (SFR) and gas-phase metallicity ($Z_g$). It has been found that at fixed stellar mass ($M_\star$), galaxies with higher SFR  show lower $Z_g$ \citep{2008ApJ...672L.107E,2010MNRAS.408.2115M,2010A&A...521L..53L}, which is the so-called fundamental metallicity relation (FMR). \citet{2008ApJ...672L.107E} and \citet{2010MNRAS.408.2115M} suggest that the existence of this observed anti-correlation between metallicity and SFR is qualitatively consistent with a scenario in which stochastic metal-poor gas accretion fuels star formation. In this way, the accretion of metal-poor gas does not change $M_{\star}$ much, but triggers star formation while diluting the gas, thus decreasing its mean metallicity. Eventually, star formation consumes the gas and stellar winds and supernova ejecta increase the gas metallicity, until new metal-poor gas is accreted and the process starts over. The explanation of the FMR within this scenario has been also probed quantitatively by several recent numerical simulations and analytic models  \citep[for more details, see][]{2017ASSL..430...67S,2014A&ARv..22...71S}. There is also a local FMR, i.e., observational evidence for a local enhancement in SFR coinciding with metallicity drops in spatially resolved  galaxies \citep{2010Natur.467..811C,2013ApJ...767...74S,2015ApJ...810L..15S}. Chemical inhomogeneities have been found across the discs of star-forming galaxies in the local Universe, where regions with low gas-phase metallicity often coincide with enhanced SFR \citep{2018MNRAS.476.4765S,2019ApJ...872..144H,2019ApJ...882....9S}. In particular, \citet{2019ApJ...882....9S} found a local relation between  $Z_g$ and SFR surface density studying a large number of nearby star-forming spiral galaxies from the MaNGA survey \citep{2015ApJ...798....7B}. Metal-poor galaxies at the low-mass end exhibit an anti-correlation, whereas the trend reverses for more metal-rich systems at the high-mass end. Yet again, this result has been argued as evidence of gas accretion regulated by the gravitational potential of the galaxy.

  Over the last few decades,  cosmological numerical simulations have become powerful tools for developing insight on how galaxies form and evolve  \citep[for an updated review, see][]{2020NatRP...2...42V}. Recent simulations are able to generate realistic galaxies which reproduce a diverseness of observables, mainly due to the advance in understanding the physical processes that drive the growth and evolution of galaxies, the improvement of numerical methods, and the increase of computing power. In particular, cosmological hydrodynamical simulations have made remarkable progress in recent years, both large-volume simulations, e.g., \textsc{eagle} \citep{2015MNRAS.446..521S, 2015MNRAS.450.1937C}, \textsc{IllustrisTNG} \citep{2018MNRAS.473.4077P} or \textsc{horizon-agn} \citep{2014MNRAS.444.1453D}, that provide statistical samples of galaxies, and zoom-in simulations, e.g., \textsc{vela} \citep{2014MNRAS.442.1545C,2015MNRAS.450.2327Z}, \textsc{nihao} \citep{wang2015nihao}, \textsc{aspostle} \citep{sawala2016apostle} or \textsc{latte/fire} \citep{wetzel2016reconciling}, that render the small-scale view. Current simulations can reasonably reproduce galaxy populations, galaxy properties, and several observed scaling relations that were not used in their calibration. 

Understanding the history of the gas fuelling star formation in these simulations is fundamental. Since the need for gas accretion has been evidenced in simulations, they are the reference to assess the reliability of the supposed observational hints for accretion. Thus, there is growing number of  works dissecting simulations to probe specific aspects of the process;  e.g.,
    \citet{2016MNRAS.457.2605C} show a few high redshift galaxies whose discs receive external metal-poor gas clumps that spiral in and form star clusters.
    \citet{2017MNRAS.470.4698A} study the baryon cycle in the  {\sc fire} simulation concluding that accretion dominates the early growth of galaxies of all masses, while the re-accretion of gas previously ejected in galactic winds often becomes important latter on. Wind recycling occurs at the scale of the CGM. \citet{2019ApJ...875...54H} carry out a pilot study based on some 15 {\sc eagle} galaxies at low redshift  showing how the cold gas is accreted anisotropically. It is concentrated and co-rotates in the plane of the galaxy disk, and could be detected in quasar sightlines up to 60 kpc from the galaxy center.
    \citet{2020MNRAS.497.4495M} study inflows and wind recycling in the {\sc eagle} simulation, to find that the majority of the accreted gas is infalling for the first time.
    \citet{2020MNRAS.495.2827C} analyze the relation between the gas accretion rate and the radial variation of $Z_g$, finding a clear relation  where  larger rates are associated with steeper slopes.
\citet{2021arXiv210210913W} use the {\sc eagle} simulations to characterize the gas accreting onto haloes, its  density, temperature, and metallicity, classifying accretion as hot or cold based on temperature. On first-infall onto a halo, the accreting gas has metallicity one order of magnitude smaller than the CGM and two orders of magnitude smaller than the ISM.

Here, we continue this line of studies focusing on two other particular aspects related to understanding the observed relation between metallicity and enhanced star-formation. Specifically, we  investigate whether the observed local relation between $Z_g$ and SFR is reproduced by the simulations, and to what degree metal-poor gas accretion drives star formation in the local Universe.  
We employ  the high resolution simulation of the \textsc{eagle} suite (details are given in Section~\ref{sec:eagle}). The \textsc{eagle} simulations reproduce a large number of observations, including the evolution of the galaxy stellar mass function \citep{2015MNRAS.450.4486F}, the evolution of galaxy sizes \citep{2017MNRAS.465..722F} and colors \citep{2015MNRAS.452.2879T}. They are also able to reproduce scaling relations, such as the correlation between stellar mass and gas-phase
metallicity at redshift~zero \citep{2015MNRAS.446..521S}, the FMR up to redshift~5 \citep{2017MNRAS.472.3354D} and the relation between metallicity and galaxy size, at fixed stellar mass \citep{ 2018ApJ...859..109S}. \citet{2016MNRAS.459.2632D} also found a relation between neutral gas fractions, stellar masses, and present SFRs. While these scaling relations are formulated in terms of integrated galaxy quantities, the study of local relations between 
physical properties in galaxies can provide further insight into galaxy evolution. For instance, \citet{2019MNRAS.485.5715T} also found the stellar mass-metallicity relation locally, employing spatially resolved regions in galaxies. 

The paper is organized as follows: Section~\ref{sec:eagle} gives an overview of the \textsc{eagle} simulation suite and its database. Section~\ref{section:dataanalysis} describes the selection of galaxies used in our study, specifically,  Section~\ref{sec:disk}  refers to disc galaxies and the synthesis of maps of physical quantities (Section~\ref{sect:maps}) whereas Section~\ref{sec:sampleselec} describes the sample of central galaxies used to trace back in time (Section~\ref{subsection:partanalysis}) the origin of the gas. The maps of metallicity and star formation are compared with MaNGA observations in Section~\ref{res:2}. The origin of the gas for each stellar mass is dissected and quantified in Section~\ref{res:1}, including its metallicity (Section~\ref{sec:metallicity}). Finally, the results are discussed in Section~\ref{sect:conclusions}.

\section{The EAGLE simulation}
\label{sec:eagle}
We make use of the suite of cosmological hydrodynamic simulations of the Evolution and Assembly of GaLaxies and their Environments (\textsc{eagle}) project \citep{2015MNRAS.446..521S, 2015MNRAS.450.1937C}. These simulations were performed using a modified version of the N-body Tree-Particle-Mesh (TreePM) smoothed particle hydrodynamics (SPH) code \textsc{gadget}-3, which is based on the \textsc{gadget}-2 code, last described by \citet{2005MNRAS.364.1105S}. The main modifications of the code are the update of the formulation of the SPH  \citep{2015MNRAS.454.2277S}, the time-stepping criteria \citep{2012MNRAS.419..465D}, and the implementation of a large number of subgrid routines that account for physical processes occurring on scales below the resolution limit of the simulations, such as radiative cooling \citep{2009MNRAS.393...99W}, star formation  \citep{2008MNRAS.383.1210S}, stellar evolution and metal enrichment \citep{2009MNRAS.399..574W}, black hole growth \citep{2005Natur.435..629S,2009MNRAS.398...53B,2015MNRAS.454.1038R}, and feedback from stars \citep{2012MNRAS.426..140D}. These subgrid recipes are systematically described by \citet{2015MNRAS.450.1937C} and \citet{2015MNRAS.446..521S}, and we refer the reader to these papers. However, since our study heavily relies on metallicities and star formation rates, the corresponding sub-grid prescriptions will be outlined for completeness.

The numerical implementation of star formation in {\sc eagle} is an extension of the model of \citet{2008MNRAS.383.1210S}. The stochastic conversion of gas particles into single stellar population (SSP) star particles is based on the volumetric conversion of the Kennicut-Schmidt law \citep{1998ApJ...498..541K}, and function of the thermodynamical state of the medium, namely its pressure. Above some resolution criterion, an effective, polytropic equation of state is imposed \citep[with polytropic index 4/3, chosen to prevent numerical fragmentation;][]{2008MNRAS.383.1210S}. In addition to the model of  \citet{2008MNRAS.383.1210S}, the EAGLE implementation includes a metallicity-dependent threshold density for star formation \citep{2004ApJ...609..667S}. This is motivate by the fact that metal cooling accelerates the transition from a warm (neutral) to a cold (molecular) phase, allowing for star formation at initially lower gas density. The probability for a gas particle to be converted into a star particle is calculated for particles with density above the threshold and temperature within some specified range.

  The mass (and energy) released by evolving single stellar populations (SSPs) is deposited into the gas phase at each simulation time step, following the model of \citet{2009MNRAS.399..574W}. The SSPs are described by a \citet{2003ApJ...586L.133C} initial mass function (IMF) ranging from $0.1$ to $100~\mathrm{M}_\odot$, and are evolved according to the metallicity-dependent lifetimes of \citet{1998A&A...334..505P}. The latter gives the range of stellar masses that reach the end of their main-sequence phase within the simulation time step. The total mass and element masses ejected by the SSP are computed combining the star particle mass fraction reaching the aforementioned evolutionary stage, its initial element abundance and the nucleosynthetic yields for asymptotic giant branch stars, massive stars, and core-collapse supernovae \citep[from][]{2001A&A...370..194M,1998A&A...334..505P}. The mass lost from Type Ia supernovae is also taken into account.  The ejected mass is distributed to neighbouring gas particles with weights that are function of the SPH kernel size and the distance from the star particle. The metals C, N, O, Ne, Mg, Si, Ca, S, and Fe are tracked individually, and their total mass fraction provides the metallicity.

The metallicities thus computed depend on the adopted stellar yields and IMF, and this assumption leads to a bias in the predicted metallicities of the order of a few times \citep[e.g.,][]{2009MNRAS.399..574W}. Similar or even larger biases are also present when measuring metallicities \citep[e.g.,][]{2008ApJ...681.1183K}. The existence of these systematic errors cautions against a trivial direct comparison between observed and predicted metallicities \citep[][]{2015MNRAS.446..521S}. However, relative variations are much less subject to systematics and, in this case, {\sc eagle} predictions have been proven to provide fair quantitative agreement with observations \citep[e.g.,][]{2018ApJ...859..109S}.

The \textsc{eagle} simulations adopt a flat $\Lambda$CDM cosmology with parameters derived from the Planck mission data \citep{2014A&A...571A...1P} ($\Omega_m$ = 0.307, $\Omega_{\Lambda}$ = 0.693, $\Omega_b$ = 0.04825, $h$ = 0.6777, $\sigma_8$ = 0.8288, $n_s$ = 0.9611, $Y$ = 0.248) and track the evolution of baryonic and non-baryonic matter from $z = 127$ to present day. 

The \textsc{eagle} suite consists of independent simulations or models with varying box sizes and resolutions. The subgrid recipes of the \textsc{reference} model are calibrated to reproduce a limited subset of observations of galaxies at $z\sim0$, namely the galaxy stellar mass function, galaxy sizes, and the stellar mass--black hole mass relation. {In this work we use the \textsc{recal}L0025N0752 simulation, whose model is tuned to meet the calibration criteria of the \textsc{reference} model, at higher resolution.  \textsc{recal}L0025N0752 has a volume of side 25 cMpc, that initially contains  $2 \times 752^3$ dark matter and gas particles. The baryon (dark matter) particle mass is $m_\mathrm{g}=2.26 \times 10^5~\mathrm{M}_{\odot}$ ($m_\mathrm{dm}=1.21 \times 10^6~\mathrm{M}_{\odot}$), and the maximum gravitational softening length is $\epsilon_\mathrm{prop}=0.35~\mathrm{pkpc}$. \textsc{recal}L0025N0752 was chosen because it provides the highest resolution among those in the {\sc eagle} suite. 

The EAGLE database \citep{2016A&C....15...72M} provides a considerable amount of information on structures identified within the simulation volume: overdense regions selected with the friends-of-friends \citep[FoF, ][]{1985ApJ...292..371D} algorithm and the substructures in them, identified as self-bound haloes with the SUBFIND algorithm \citep{2001MNRAS.328..726S, 2009MNRAS.399..497D}.
Along with integrated galaxy/halo physical quantities, the database provides also particle's raw data in the form of 29 simulation snapshots distributed in time between $z=20$ and $z=0$ \citep{2017arXiv170609899T}.

\subsection{Data sample selection and analysis}
\label{section:dataanalysis}

We carry out two analyses in this work. The selection of the galaxies is detailed in  Sections~\ref{sec:disk} and \ref{sec:sampleselec}.
In the first analysis, we spatially resolve the gas-phase metallicity  and the star formation rate surface density ($\Sigma_{\rm SFR}$) of star-forming galaxies. In order to compare them with the observed local $Z_g$--$\Sigma_{\rm SFR}$ relation, we select disc galaxies at $z=0$. The construction of maps of physical quantities is detailed in Section~\ref{sect:maps}.
In the second analysis, we study the source of the gas that sustains star formation. In order to do so, we trace back in time the gas that recently formed stars in the galaxies, as described in Section~\ref{subsection:partanalysis}.
The galaxies selected for studying the  local $Z_g$--$\Sigma_{\rm SFR}$ relation are a subset of the galaxies used in the analysis of the source of gas.

\subsubsection{Selection of central disc galaxies}
\label{sec:disk}
The first part of the work requires a sample of central galaxies at $z=0$ that have {\em disc-like} features. We select galaxies that have a significant fraction of kinetic energy in ordered co-rotation, are oblate, and relatively flat. Details will be given below, but all in all, we obtain a subsample of 107 {\em disky} central galaxies using these constrains, with their mass distribution included in Figure \ref{fig:samples} as a blue dashed line.
\begin{figure}
    \centering
    \includegraphics[scale = 0.4]{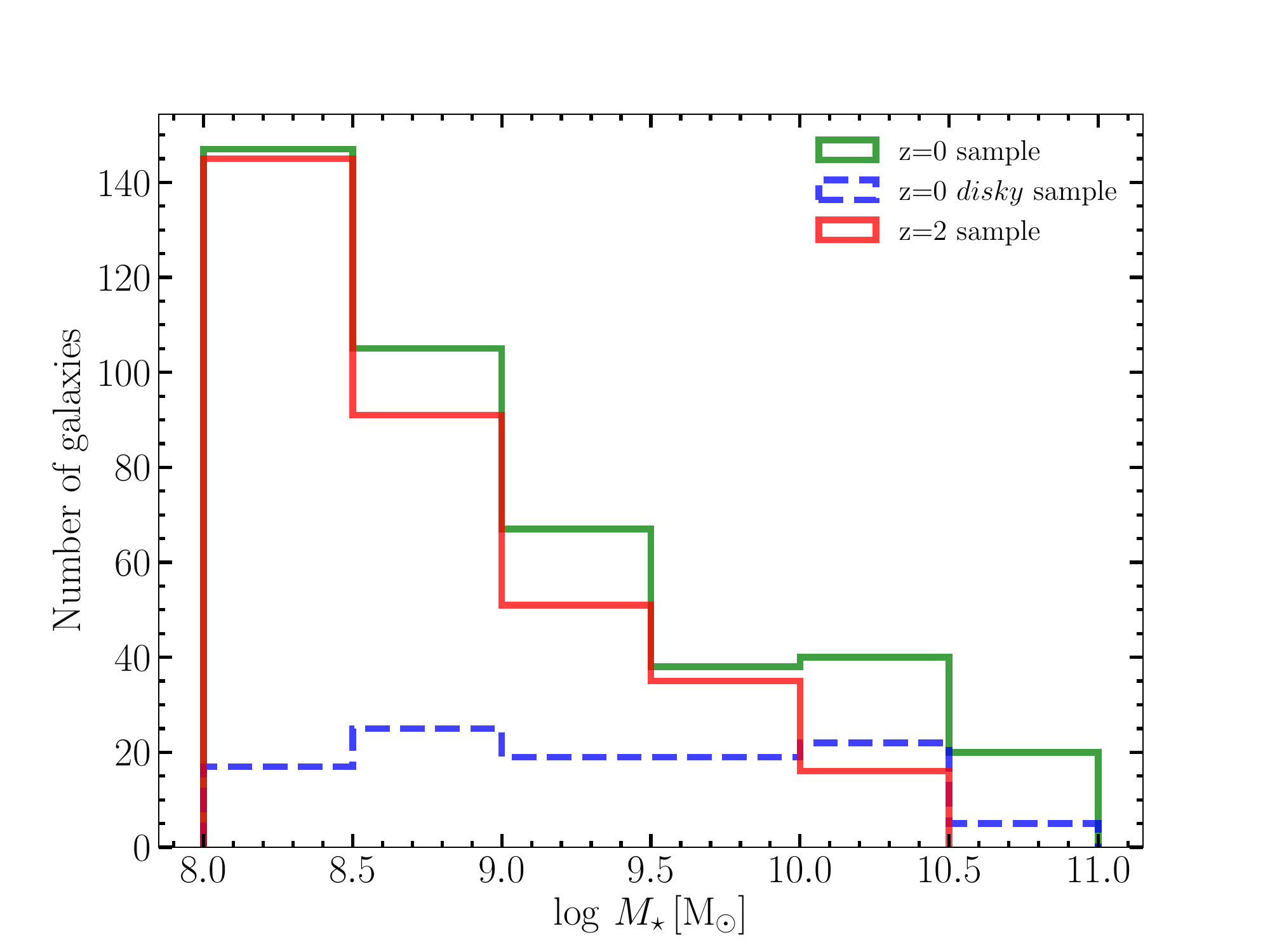}
    \caption{Number of selected galaxies as a function of stellar mass. The green and red solid lines correspond to the samples of central galaxies at $z=0$ and $z=2$, respectively. The blue dashed line corresponds to the subsample of central galaxies that have {\em disc-like} features at $z=0$.
    }
    \label{fig:samples}
\end{figure}

In order to identify galaxies with {\em disky} morphology, we follow} the approach described by \citet{2017MNRAS.472L..45C}, who
found that the morphology of \textsc{eagle} galaxies can be characterised through their rotational kinetic energy. Galaxies with $\kappa_\mathrm{co}>0.4$ correspond to {\em disky} galaxies, with $\kappa_\mathrm{co}$ the fraction of kinetic energy in ordered co-rotation,
\begin{equation}
\kappa_{\mathrm{co}}=\frac{K_{\mathrm{co}}^{\mathrm{rot}}}{K},
   \label{eq:Kco}
\end{equation}
where $K$ is the total kinetic energy,
\begin{equation}
K=\frac{1}{2} \sum_{i}  m_{i} v_{i}^{2},
  \label{eq:Kco1}
\end{equation}
and $K^\mathrm{rot}_\mathrm{co}$ represents the total kinetic energy in ordered co-rotation,
\begin{equation}
K_\mathrm{co}^\mathrm{rot}=
\frac{1}{2} \sum_{L_{i}>0}  m_{i}\left(\frac{L_{i}}{m_{i} R_{i}}\right)^{2}.
  \label{eq:Kco2}
\end{equation}
The symbols in the above equations stand for: the mass of each stellar particle $m_i$,  the velocity with respect to the center of mass $v_i$, the angular momentum along the direction of the total angular momentum of the stellar component of the galaxy $L_{i}$, and  the projected distance to the rotation axis $R_i$. The sum in Eq.~(\ref{eq:Kco1}) goes through all stellar particles within 30 pkpc whereas the sum in Eq.~(\ref{eq:Kco2}) only includes co-rotating $\left(L_{i}>0\right)$ stellar particles.
To be in the safe side, we also impose constrains on the galaxy shape parameters as described by \citet{2019MNRAS.485..972T}, who characterised the stellar morphologies of \textsc{eagle} galaxies by modelling the spatial distribution of their stars with an ellipsoid described by the flattening ($\epsilon$) and triaxiality ($T$) parameters. These parameters are defined as,
\begin{equation}
    \epsilon=1-\frac{c}{a}, \text { ~~~~~~and~~~~~~~ } T=\frac{a^{2}-b^{2}}{a^{2}-c^{2}},
\end{equation}
where $a$, $b$, and $c$ are the moduli of the major, intermediate, and
minor axes of the ellipsoid, respectively. They impose $T<0.3$ for the galaxies to be oblate, and $\epsilon > 0.4$ to be flat, both characteristics of disky galaxies.

\begin{figure}
    \centering
    \includegraphics[scale = 0.42]{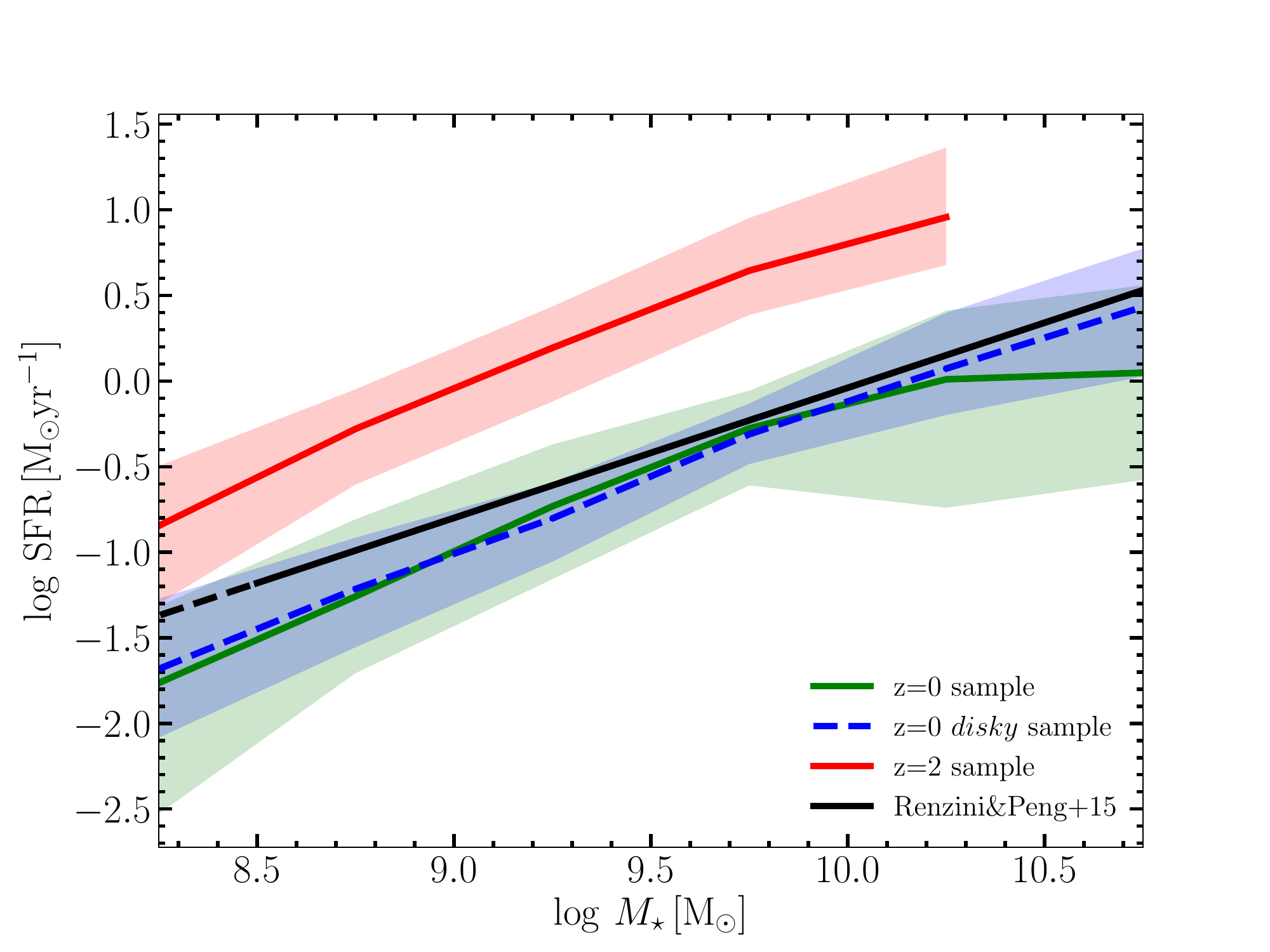}
    \caption{Star-formation main sequence of the selected galaxies, i.e., SFR versus $M_\star$ in log-spaced stellar mass bins of 0.5 dex. The blue dashed line corresponds to the median SFR of the sample of disc galaxies at $z=0$. The green and red solid lines show  the median SFR of the samples at $z=0$, and $z=2$, respectively. The colored regions indicate the $\mathrm{10^{th}}$ and $\mathrm{90^{th}}$ percentiles.  The black solid line corresponds to the best straight-line fit to the main sequence of local galaxies in the SDSS database \citep{2015ApJ...801L..29R}, with the black dashed line showing its linear extrapolation to low mass.
    }     
    \label{fig:mainseq}
\end{figure}

The \textsc{eagle} database provides $\kappa_{\rm co}$, $T$, and $\epsilon$ for individual galaxies, and we select all $z=0$ central galaxies  that have $\kappa_{\rm co}>0.4$,  are oblate ($T<0.3$) and flat ($\epsilon > 0.4$), which renders our sample of 107 disky galaxies. They have a rather uniform mass distribution (blue dashed line in Figure~\ref{fig:samples}) and are in the star-formation main sequence (blue dashed in Figure~\ref{fig:mainseq}). 
We note the small number of disc galaxies with low mass compared to the full sample of central galaxies at $z=0$ (Figure \ref{fig:samples}, the green line, to be discussed in Sect.~\ref{sec:sampleselec}). This reduction is due to our morphological selection criteria and the constraint based on computing kinetic energy in co-rotation. Owing to the resolution limit of the simulation, low-mass galaxies often do not have enough stellar particles to be classified and, consequently, are not selected as discs.

\subsubsection{Maps of\, $Z_g$ and $\mathrm{\Sigma_{SFR}}$ of the star-forming gas }\label{sect:maps}
In order to analyze the spatially resolved relation between star formation rate and gas-phase metallicity in the simulation, we synthesized maps of $\Sigma_{\rm SFR}$ and $Z_g$ for the sample of disc galaxies selected in Section~\ref{sec:disk}. We set them in a face-on configuration using their spin per unit mass, which is obtained from the \textsc{eagle} database.  Properties are projected in the plane perpendicular this spin as follows: for each galaxy, we select all the star-forming gas particles at the same projected distance from the center of the galaxy, defined as its center of potential. For those particles, we calculate their star formation rate densities and mass-weighted average metallicities in $1~\mathrm{kpc^2}$ areas. For this we use the individual gas-particle SFR and $Z_g$, which are available in the \textsc{eagle} snapshots.
 
\subsubsection{Selection of central galaxies at $z=0$\, and $z=2$}
\label{sec:sampleselec}

We lift the constraints used in Section~\ref{sec:disk} to improve the statistics.  
All \textsc{eagle} central galaxies\footnote{The galaxy located at the minimum of the potential of a FoF group is defined as central; all others are named satellites.} satisfying $M_\star\geq 10^8$ $\mathrm{M_{\odot}}$ are included in this second sample. The stellar mass threshold minimizes any potential bias induced by the finite resolution of the simulation, selecting galaxies with enough particles for our analysis to be reliable. The selection was carried out at two different redshifts, $z=0$ and $z=2$ (actually $z=2.01$, but labelled $2$ throughout the paper). Redshift zero was selected to have a specific prediction for the local galaxies whereas redshift 2 portrays  the maximum star formation rate density of the Universe, which then declines exponentially at later times \citep[e.g.,][]{2014ARA&A..52..415M}.\footnote{Incidentally, we note that the shape of this decline with redshift in \textsc{eagle} is in excellent agreement with observations \citep{2015MNRAS.450.4486F}.} We also impose an upper limit on the stellar mass, excluding galaxies above $10^{11}$ $\mathrm{M_{\odot}}$ at $z=0$ and above $10^{10.5}$ $\mathrm{M_{\odot}}$  at $z=2$ since only one central galaxy has mass above each one of these limits.

Thus, our sample at $z=0$ consists of 417 central galaxies with $M_\star$ ranging from $10^8$ to $10^{11}\ \mathrm{M_{\odot}}$, and analogously, the sample at $z=2$ comprises 338 central galaxies with $M_\star$ between $10^8$ to $10^{10.5} \ \mathrm{M_{\odot}}$. The solid lines in Figure \ref{fig:samples} show the mass distribution of the samples at $z=0$ (green) and $z=2$ (red) in log-spaced stellar mass bins of 0.5~dex. Low-mass galaxies outnumber both samples as galaxies with $M_\star<10^{9} \ \mathrm{M_{\odot}}$ correspond to 60.4\% and 69.8\% of the galaxies at $z=0$ and $z=2$, respectively.
Note that, by construction, the set of 107 disc galaxies selected in Section~\ref{sec:disk} is a subset of the 417 $z=0$ galaxies.

The selected galaxies portray regular star-forming galaxies. The star-forming galaxy main sequence of the selected galaxies is shown in Figure \ref{fig:mainseq}. We represent the galaxy SFR in log-spaced stellar mass bins of 0.5 dex. The red and green solid lines show the median of the star formation rate of the galaxies in the samples at $z=0$ and $z=2$, respectively. The colored region indicates the $\mathrm{10^{th}}$ and $\mathrm{90^{th}}$ percentiles. The black solid line corresponds to the best straight-line fit to the main sequence of local galaxies in the SDSS database \citep{2015ApJ...801L..29R}. It was included for reference to show that the star-forming main sequence of the galaxies selected at $z=0$ is in agreement with observations.

%
\begin{table*}
\small
\centering
\begin{tabular}{|c|c|c|l|}
\hline

\multicolumn{3}{|c|}{CATEGORIES}                                                                                                                                                  &        ~~~~~DEFINITION                                                                                                                           \\ \hline \hline
\multicolumn{3}{|c|}{ \footnotesize{INSIDE THE GALAXY}}                                                                                                                                           & \begin{tabular}[c]{@{}l@{}}Gas that was already gravitationally bound \\to the galaxy at the previous redshift\end{tabular}                            \\ \hline
\multirow{4}{*}{\begin{tabular}[c]{@{}c@{}} \\~\\\footnotesize{OUTSIDE}\\ \footnotesize{THE}\\ \footnotesize{GALAXY}\end{tabular}} & \multicolumn{2}{c|}{\footnotesize{UNBOUND}}                                                                    & \begin{tabular}[c]{@{}l@{}}Gas that was not gravitationally bound to\\  any subhalo at the previous redshift\end{tabular}                           \\ \cline{2-4} 
                                                                                & \multirow{3}{*}{\begin{tabular}[c]{@{}c@{}}\\\footnotesize{IN} \\ \footnotesize{ANOTHER} \\ \footnotesize{GALAXY}\end{tabular}} & \footnotesize{DARK GALAXY}  & \begin{tabular}[c]{@{}l@{}}Gas bound to a dark galaxy,  which is a\\ subhalo mainly formed by gas and dark matter,\\ with $M_\star <$~5\% of the baryon mass \end{tabular} \\ \cline{3-4} 
                                                                                &                                                                                  & \footnotesize{MINOR MERGER} & \begin{tabular}[c]{@{}l@{}}Gas bound to a subhalo that produced a \\ minor merger (mass ratios $< 1/3$)\end{tabular}                          \\ \cline{3-4} 
                                                                                &                                                                                  & \footnotesize{MAJOR MERGER} & \begin{tabular}[c]{@{}l@{}}Gas bound to a subhalo that produced a \\ major merger (mass ratios $\geq 1/3$)\end{tabular}                       \\ \hline
\end{tabular}
\caption{
Classification of gas particles according to their location in the redshift previous to form stars.}
\label{tab:cat}
\end{table*}
\begin{figure*}
    \centering
   \includegraphics[width=0.9\textwidth]{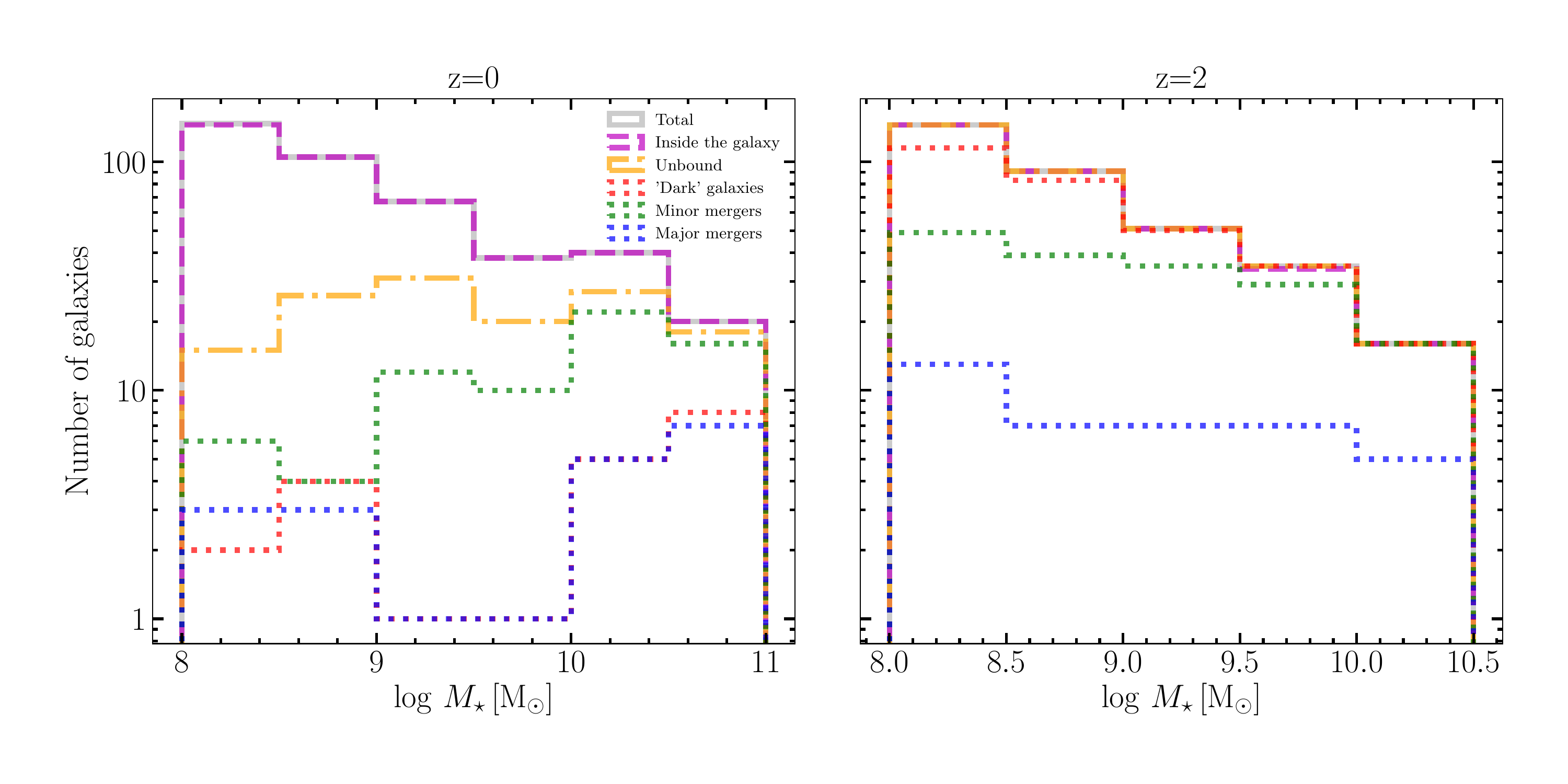}
    \caption{Number of central galaxies with stellar mass $M_\star$ that receive gas from each category. Left panel: centrals at $z=0$ with the new stars formed between $z=0.1$ and $z=0$, i.e., new stars younger than 1.34 Gyr. Right panel: centrals at $z=2$ with the new stars formed between $z=3$ and $z=2$, i.e., new stars younger than 1.13 Gyr.  As indicated in the insets, the different lines and colors represent the different origins of the star-forming gas; see Table~\ref{tab:cat}. The grey line corresponds to the total number of galaxies. Note that galaxies contribute to more than one category if the gas they employ to form new stars has several origins. 
    }
    \label{fig:num}
\end{figure*}
\subsubsection{Tracking back in time the gas that recently formed stars}
\label{subsection:partanalysis}
In order to track the gas that recently formed new stars in the galaxies,  we select the star particles that are formed between two snapshots, thus in the redshift interval $(z_\mathrm{prev},z)$, where $z < z_\mathrm{prev}$. By construction, these stars were gas particles at  $z_\mathrm{prev}$, and {\sc eagle} makes it simple to match  baryonic particles through time because they do not change their identifier during the simulation, even when gas turns into stars.  The gas that formed new stars was classified within one of the following five classes:
\begin{itemize}
\item[-] the gas was already gravitationally bound to the galaxy at $z_\mathrm{prev}$;
\item[-] the gas was not bound to any resolved subhalo at $z_\mathrm{prev}$;
\item[-] the gas was bound to a subhalo at $z_\mathrm{prev}$ that produced a minor merger with the galaxy;
\item[-] the gas was bound to a subhalo at $z_\mathrm{prev}$ that produced a major merger with the galaxy;
\item[-] the gas was bound to a {\em dark} galaxy at $z_\mathrm{prev}$ that merged with the galaxy.
\end{itemize}
%
%
%
%
%
%
%
%
We define as dark galaxy any subhalo in which the stellar mass is less than five percent of its baryonic mass, i.e., galaxies primarily formed by gas and dark matter. Minor and major mergers are defined by the stellar mass ratio of the merging galaxies, $< 1/3$ and $ \geq 1/3$, respectively. These five categories are named and described in Table~\ref{tab:cat}.

We carry out the classification at two different epochs. At $z=0$, we study stars formed between $z_\mathrm{prev}=0.1$ and $z=0$, i.e, stars younger than $1.34~\mathrm{Gyr}$. At $z=2$, we select $z_\mathrm{prev}=3$ (actually,  3.02) so that the time interval between  shapshots is similar in both analyses (1.13 Gyr in this case). We note that these time intervals are typically larger than several time-scales relevant to the star-formation process. The gas consumption time-scale varies from 0.5 to 2 Gyr for galaxies at redshifts between 2 and 0, and it gets reduced by up to one order of magnitude if galaxies have intense winds \citep[e.g.,][]{2014A&ARv..22...71S}. The free-fall time scales with the mean density of the system, and is similar to the dynamical time \citep[e.g., the rotational period around the centre of the galaxy;][]{2008gady.book.....B}. For the Milky Way, both time-scales are of the order 0.25 Gyr, and they decrease with decreasing galaxy mass. Galaxy winds are common. With speeds of 100 km\,s$^{-1}$ or more, they eject materials out of the star-forming regions which set a time-scale for this material to reach the CGM of the order of 0.2 Gyr \citep[e.g.,][]{2017ApJ...834..181O}. All these physical processes are encompassed and averaged out by our analysis.

Figure~\ref{fig:num} shows the number of galaxies forming new stars in each one of the five categories defined above. They are separated by stellar mass, with the left and right panels corresponding to galaxies at $z=0$ and $z=2$, respectively. Different colors distinguish the different location of the star-forming gas when it was at $z=0.1$ (left panel) and $z=3$ (right panel). 
Only two low-mass redshift-zero galaxies  did not form new stars in any category, and they were excluded from all the further analysis.
The grey histogram gives to the total number of galaxies in each mass bin. Pre-existing gas always contributes to the recent star-formation in all  galaxies. However, accreted gas gains relative importance at $z=2$.  All galaxies have contribution from unbound gas, and most high-redshift galaxies (right panel) from gas coming from mergers with dark galaxies. Mergers are generally more common at high redshift. At $z=0$, in the low and intermediate mass bins, only a few galaxies formed stars using gas coming from dark galaxies and major mergers.

%
\begin{figure*}
    \centering
    \includegraphics[width=0.8\textwidth]{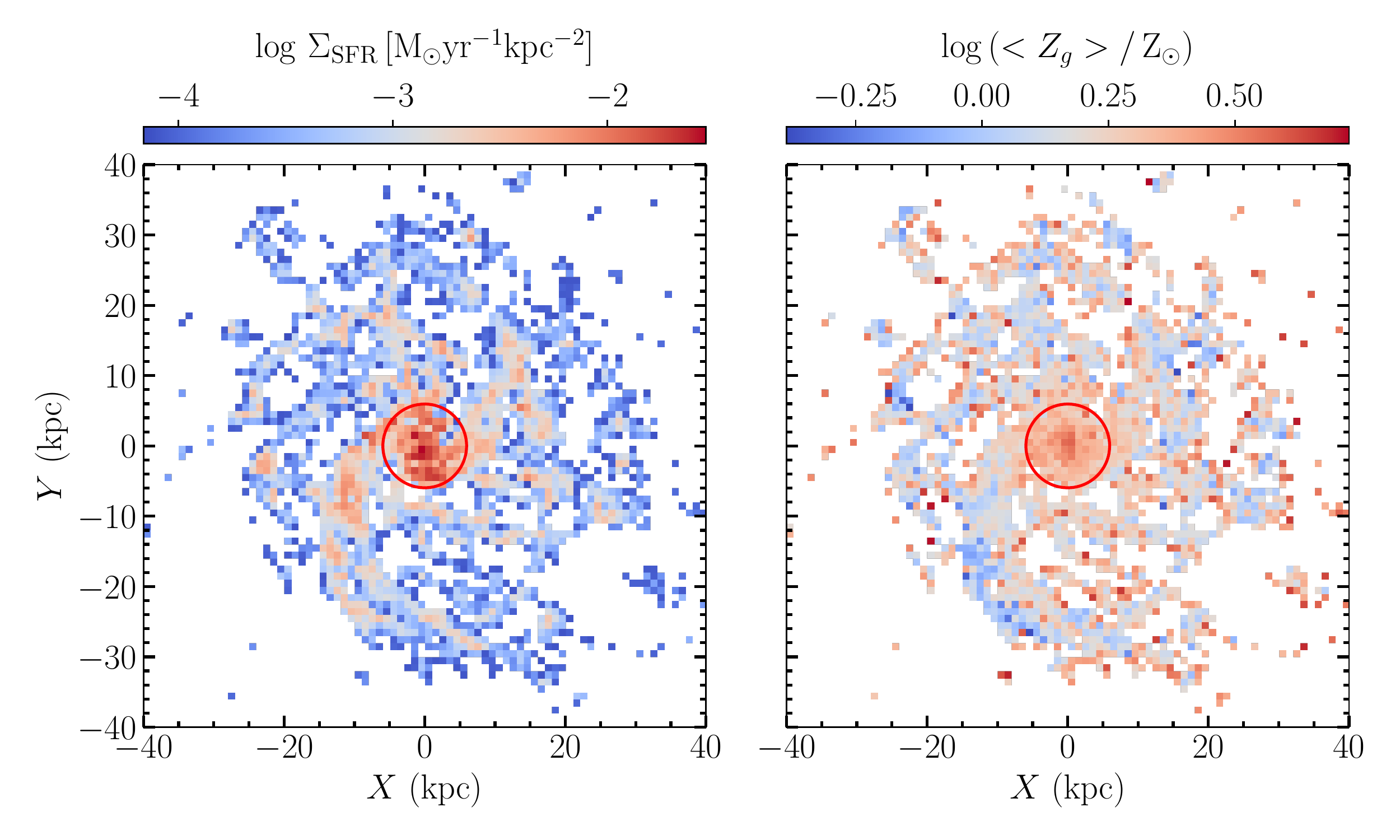}
    \caption{Star formation rate surface density (left panel) and mass-weighted average metallicity (right panel) for galaxy $j=817104$, with $\log(M_\star^j/\mathrm{M_{\odot}})=10.3$. The red circle has the radius of a sphere that encloses half of the stellar mass of the galaxy. The central increase of $\Sigma_{\rm SFR}$ comes together with an increase of  $Z_g$, meaning that the two quantities are positively correlated. 
    }
    \label{fig:g2}
\end{figure*}
\begin{figure*}
    \centering
    \includegraphics[width=0.78\textwidth]{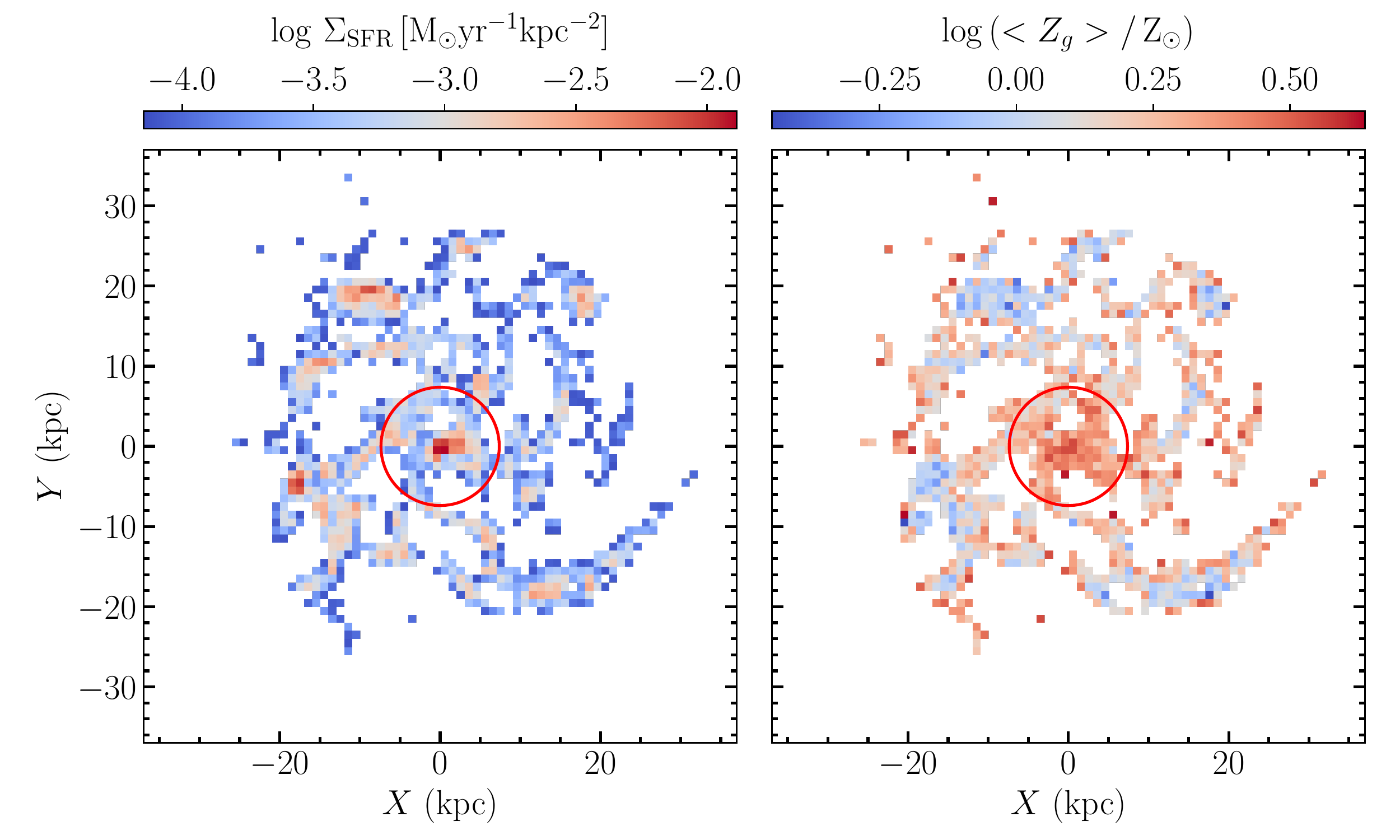}
    \caption{Star formation rate surface density (left panel) and mass-weighted average metallicity (right panel) for galaxy $j=1019529$ with $\log(M_{\star}^j/\mathrm{M_{\odot}})=9.7$. The red circle has the radius of a sphere that encloses half of the stellar mass of the galaxy.
    }
    \label{fig:g3}
\end{figure*}
\begin{figure*}
    \centering
  \includegraphics[width=0.8\textwidth]{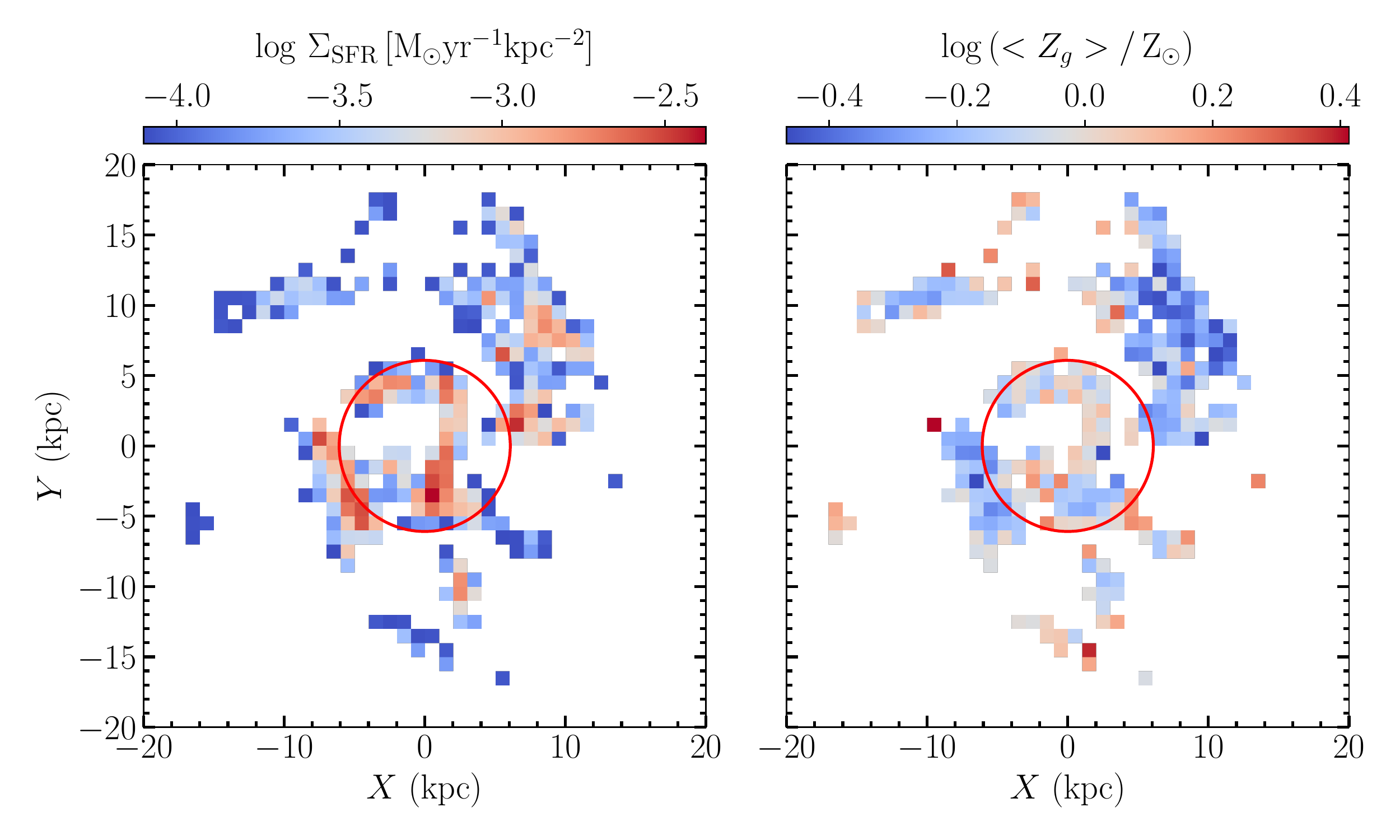}
    \caption{Star formation rate surface density (left panel) and mass-weighted average metallicity (right panel) for galaxy $j=1174755$ with $\log(M_{\star}^j/\mathrm{M_{\odot}})=8.8$. The red circle has the radius of a sphere that encloses half of the stellar mass of the galaxy. Note the clear anti-correlation between SFR surface density (left panel) and metallicity (right panel).}
    \label{fig:g4}
  \end{figure*}
\bigskip
\section{Local $Z_g-\Sigma_{\rm SFR}$ relation}\label{res:2}

The accretion process is localized, clumpy, and intermittent (Section~\ref{intro}) and it leaves imprint in the spatial distribution of gas within the discs. This distribution and the resulting spatially-resolved star-formation are studied in this section. Specifically, here we analyze a subsample of disc galaxies (see Section~\ref{sec:sampleselec}) to study whether the local $Z_g-\Sigma_{\rm SFR}$ relation seen in observations of nearby spiral galaxies \citep{2019ApJ...882....9S,2019ApJ...872..144H} is also present in the simulated disc galaxies.

 \textsc{eagle} snapshots provide physical properties for the gas particles, in particular, $Z_g$ and SFR. We use them to calculate the face-on projected SFR surface density, $\Sigma_{\rm SFR}$, and the mass-weighted average metallicity, $\langle Z_g \rangle$. The projection is made using bins of $1~\mathrm{kpc^2}$, with the galaxy in face-on configuration, and selecting star-forming gas (further details given in Section~\ref{sect:maps}). The bin size was chosen to approximately match the spatial resolution of the MaNGA galaxies analyzed by \citet{2019ApJ...882....9S}. Maps of $\Sigma_{\rm SFR}$ and $\langle Z_g\rangle$ for three representative galaxies covering the full range of stellar masses are shown in Figures~\ref{fig:g2}\,--\,\ref{fig:g4}. The red circle on the map indicates the projected radius of a sphere that encloses half of the stellar mass of the galaxy, and can be thought of as a proxy for the effective radius used by observers.

The example in Figure \ref{fig:g2} corresponds to the high-mass end of the sample, with a stellar mass of $\sim 2\times 10^{10}\ \mathrm{M_{\odot}}$. One can distinguish particular regions in the outer galaxy where higher $\Sigma_\mathrm{SFR}$ coincides with lower $Z_g$, but the correlation reverses sign in the central parts where regions of higher $\mathrm{SFR}$ also have considerably higher metallicities. Figure~\ref{fig:g3} shows a galaxy with intermediate stellar mass, $\sim 5\times 10^9\ \mathrm{M_{\odot}}$. There are regions in the outskirts where the anti-correlation between $Z_g$ and $\Sigma_\mathrm{SFR}$ is more significant than in the previous case. For galaxies at the low-mass end, as the one shown in Figure \ref{fig:g4}, with a stellar mass of $\sim 7\times 10^8\ \mathrm{M_{\odot}}$, the two quantities are clearly anti-correlated -- dark-blue regions in the left panel of Figure~\ref{fig:g4} (larger $\Sigma_\mathrm{SFR}$) are light blue in the right panel (smaller $Z_g$).
We note that \citet{2019ApJ...872..144H} analyzed a sample of nearby star-forming late-type galaxies and found that the presence of star-forming regions with anomalously low metallicites are higher in lower mass galaxies and in the outer regions of the discs, a result in excellent agreement with what was obtained in the \textsc{eagle} simulation.

\begin{figure}
\includegraphics[width=0.5\textwidth]{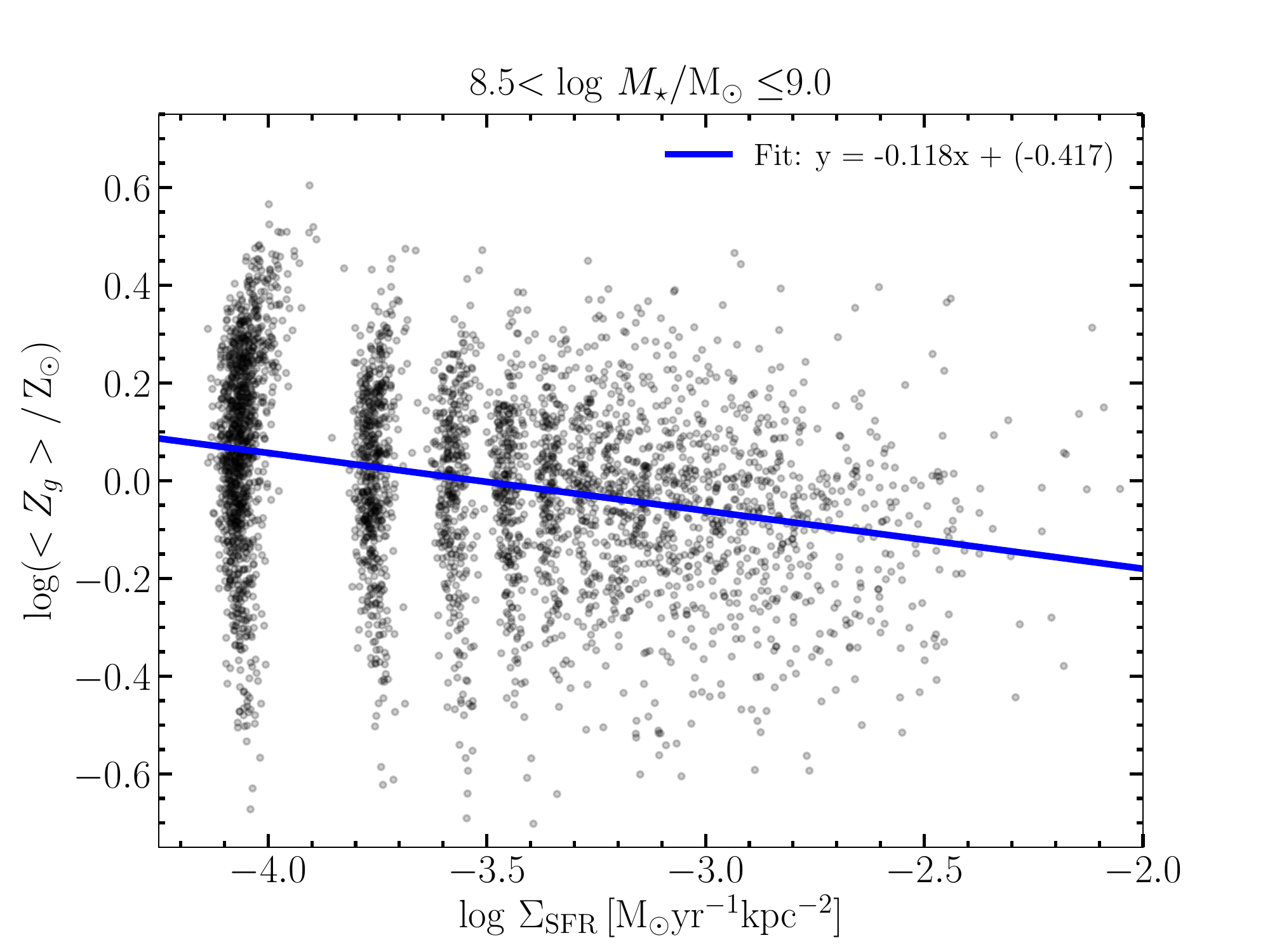}
\caption{
Scatter plot $Z_g$ versus SFR given as an example of the fits used to derive the slopes represented in Figure~\ref{fig:slopes}. It includes all galaxies in the mass bin $8.5 \leq \log(M_\star/{\rm M}_\odot) \leq 9.0$. Symbols represent individual pixels in the maps whereas the solid line indicates a straight line best fit whose parameters (slope and intercept) are given in the inset. This slope provides one of the points in Figure~\ref{fig:slopes}.
}
\label{fig:scatter}
\end{figure}

\begin{figure}
    \centering
    \includegraphics[scale = 0.4]{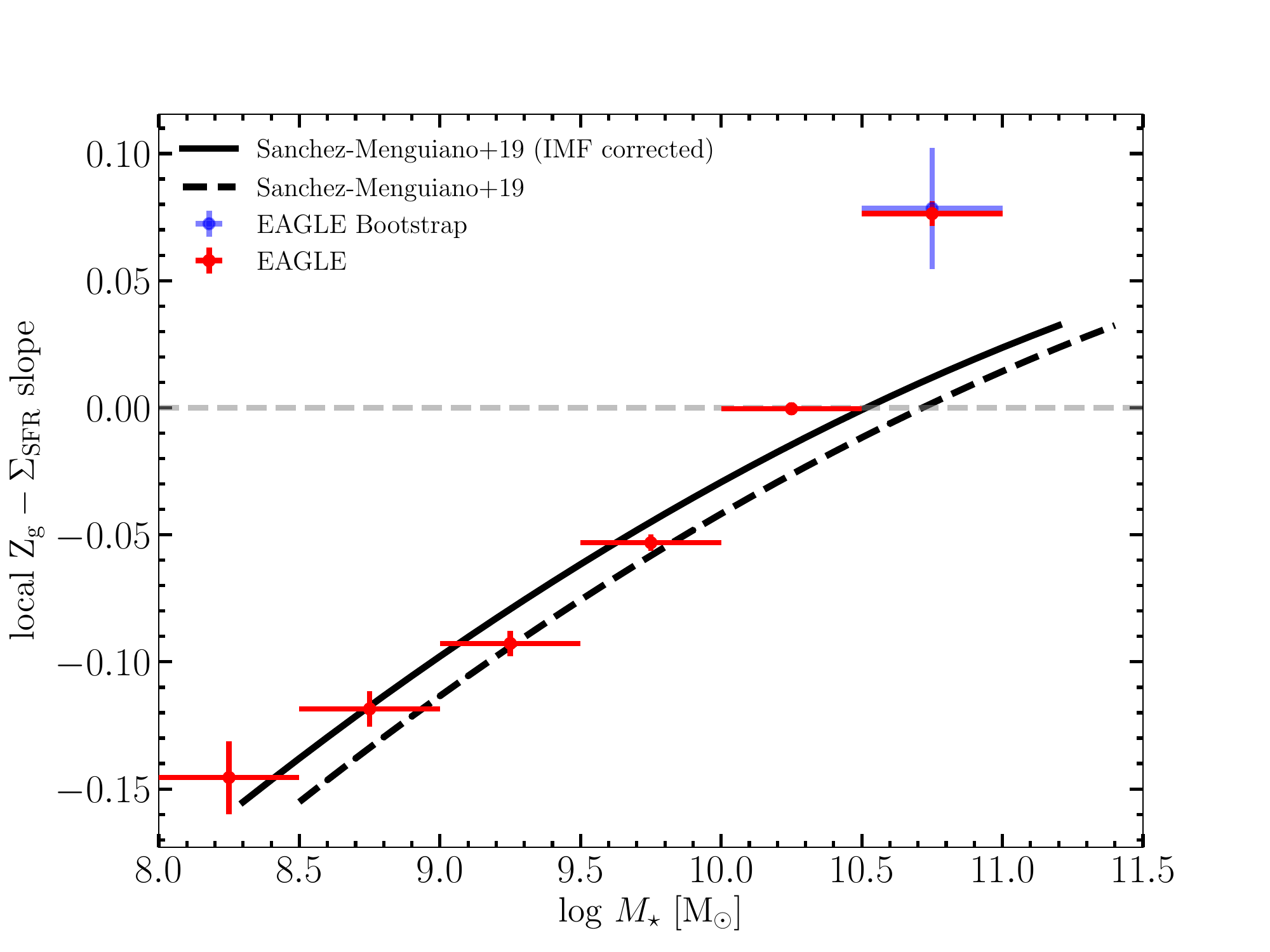}
    \caption{Local $Z_g$-- $\mathrm{\Sigma_{SFR}}$ slope as a function of galaxy stellar mass.  Negative slope means that $Z_g$ decreases when the SFR surface density increases, and vice-versa. The slopes are computed for the subsample of 107 \textsc{eagle}  central galaxies with {\em disc-like} features.  The galaxies are divided in 6 equally log-spaced stellar mass bins of 0.5 dex. For each bin, the red point corresponds to the slope of a linear fit of the mass-weighted average metallicity as a function of star formation rate surface density of all the galaxies that fall in the mass bin (see Figure~\ref{fig:scatter}).
The blue symbol represents an estimate of slope and error in the last mass bin using bootstrapping.
The black dashed line corresponds to the mean slopes observed by \citet{2019ApJ...882....9S} in a sample of 736 MaNGA galaxies. The line represents a 2nd order polynomial fit to the actual data. The solid line is the same polynomial shifted in mass to account for the different IMFs used by {\sc eagle} and MaNGA. The grey dashed line indicates no correlation and is included for reference.
   }
    \label{fig:slopes}
\end{figure}

In order to quantify the relation between $Z_g$  and $\Sigma_{\rm SFR}$, and to compare it with the relation observed by \citet{2019ApJ...882....9S}, the disc galaxy subsample was divided into six stellar mass bins of 0.5 dex.  For each mass bin, we perform linear fits to the scatter plot $\langle Z_g\rangle$ versus $\Sigma_{\rm SFR}$ i.e., a fit to the scatter plot of the two variables shown in the right and left panels of Figures \ref{fig:g2}\,--\,\ref{fig:g4}.
Figure~\ref{fig:scatter} provides as example one of the scatter plots used to infer the slopes.
Figure~\ref{fig:slopes} shows the slopes of these linear fits as a function of galaxy stellar mass. The error bars correspond to the formal errors of the linear fit \citep[obtained from the covariance matrix; e.g.,][in the y-axis]{1986nras.book.....P} and to the width of the stellar mass bin (in the x-axis). The grey dashed straight line marks slope equal to zero and is included for reference. The sign of the slope directly gives the sign of the correlation between $\Sigma_{\rm SFR}$ and $Z_g$. It is clear that the correlation is negative for low-mass galaxies and positive for high-mass galaxies, being around zero (i.e., lack of correlation) for a stellar mass $\log(M_\star/{\rm M}_\odot)\simeq 10.2$.

We compare these properties from \textsc{eagle} model galaxies with the observational results from  \citet{2019ApJ...882....9S}, who found a local relation $Z_g$--$\mathrm{\Sigma_{SFR}}$ based on 736 nearby spiral galaxies from the MaNGA survey.
The comparison is not straightforward keeping in mind the differences in the way simulated and observational relations are computed. On the one hand, observations refer to light-weighted averages inferred from the star-forming gas existing within a few effective radii from the center, to inclined galaxies, and to the residuals left when radial variations are subtracted out. On the other hand, simulations show mass-weighted averages, face-on galaxies, and use the actual maps without any cut in radial distance. Thus, we only attempt a qualitative comparison of the slopes in the simulation and in the observations to see whether \textsc{eagle} galaxies show the observed trend with galaxy stellar mass. The comparison is fair, specially, keeping in mind that making it more realistic would require carrying out a complete emission line spectral synthesis of the model galaxies, which represents a major effort clearly beyond the scope of our exploratory work.  

\citet{2019ApJ...882....9S} found the observed local $Z_g-\Sigma_{\rm SFR}$ relation to vary with stellar mass, a dependence they parameterized in terms of a 2nd order polynomial fit to the observation \citep[][Table~1]{2019ApJ...882....9S}. We include this fit as a black dashed line in Figure \ref{fig:slopes}. It is very close to the trend found in \textsc{eagle}, represented in the figure by symbols and error bars. The agreement is even better if we correct for the difference between the IMF in observations and simulations.
\citet{2019ApJ...882....9S} use the spectral-fitting code \textsc{pipe3d} \citep{sanchez2016pipe3d} to compute stellar masses, which employs single stellar population model spectra adopting a modified Salpeter IMF, whereas \textsc{eagle} uses the Chabrier IMF to produce stars \citep{2015MNRAS.446..521S}.
This leads to a systematic mass difference of about 60 percent with the \textsc{eagle} masses being smaller \citep[e.g.,][]{2001ApJ...550..212B}. The black solid line in Figure~\ref{fig:slopes} corresponds to the 2nd order polynomial fit with a shift of -0.2 dex to compensate for IMF differences.  The agreement between observations and simulations is excellent. Both show the same trend with stellar mass and a change in sign at approximately the same galaxy mass.  Below this threshold, the larger the SFR the smaller the metallicity, with the trend reversing for massive galaxies. The most significant difference is in the magnitude of the slope in the largest mass bin, with observations showing changes in metallicity smaller than numerical simulations. Such difference may be an artifact due to small number statistics since only five {\sc eagle} galaxies contribute to this mass bin. To evaluate the potential impact of having such small number of galaxies, the error of the slope was estimated by bootstrapping. Sets of 5 galaxies were randomly selected from the original sample and then processed to obtain $10^4$ slopes. The resulting mean value and standard deviation are shown as a blue symbol in Fig.~\ref{fig:slopes}. The bootstrapping based slope differs from the observed one within the error bar and so the mismatch seems to be real.

  As we mentioned above, MaNGA observations extend to only a few effective radii from the galaxy centres whereas no radial cutout has been applied to the {\sc eagle} maps. This should not make a big difference in the results since the star-forming region is within few effective radii for most of the {\sc eagle} galaxies. Moreover, there is a clear trend for negative correlations in the outskirts (Figures~\ref{fig:g2} -- \ref{fig:g4}). Excluding these regions would lead to an increase of the inferred slopes, therefore shifting upward the relation predicted by  {\sc eagle} (Figure~\ref{fig:slopes}). However, the effect should be minimal, since low-mass galaxies continue showing negative slopes even when their outskirts are removed.

The existence of regions with excess in SFR and deficit in metallicity  (Figures~\ref{fig:g2}\,--\,\ref{fig:g4}) is consistent with localized accretion of  metal-poor gas. The arrival of metal-poor gas to a particular region simultaneously triggers star formation and dilutes the pre-existing gas, thus decreasing its metallicity. This physical process has been proposed to explain the local anti-correlation observed in star-forming dwarf galaxies of the Local Universe \citep{2015ApJ...810L..15S,2018MNRAS.476.4765S} and in low-mass nearby spiral galaxies \citep{2019ApJ...882....9S}. However, it does not explain the positive correlation found in more massive galaxies (Figure~\ref{fig:slopes}).
\citet{2019ApJ...882....9S} argue that in high-mass galaxies the gas used to form new stars mostly comes from previous star-formation episodes. Thus, the regions of high gas mass, where star-formation is higher, are also relatively metal-richer. Taken together, these two mechanisms are compatible with the trend of the slope of the local $Z_g$--$\Sigma_{\rm SFR}$ relation with stellar mass portrayed in Figure~\ref{fig:slopes}.
%

%
\begin{figure*}
    \centering
    \includegraphics[width=0.9\textwidth]{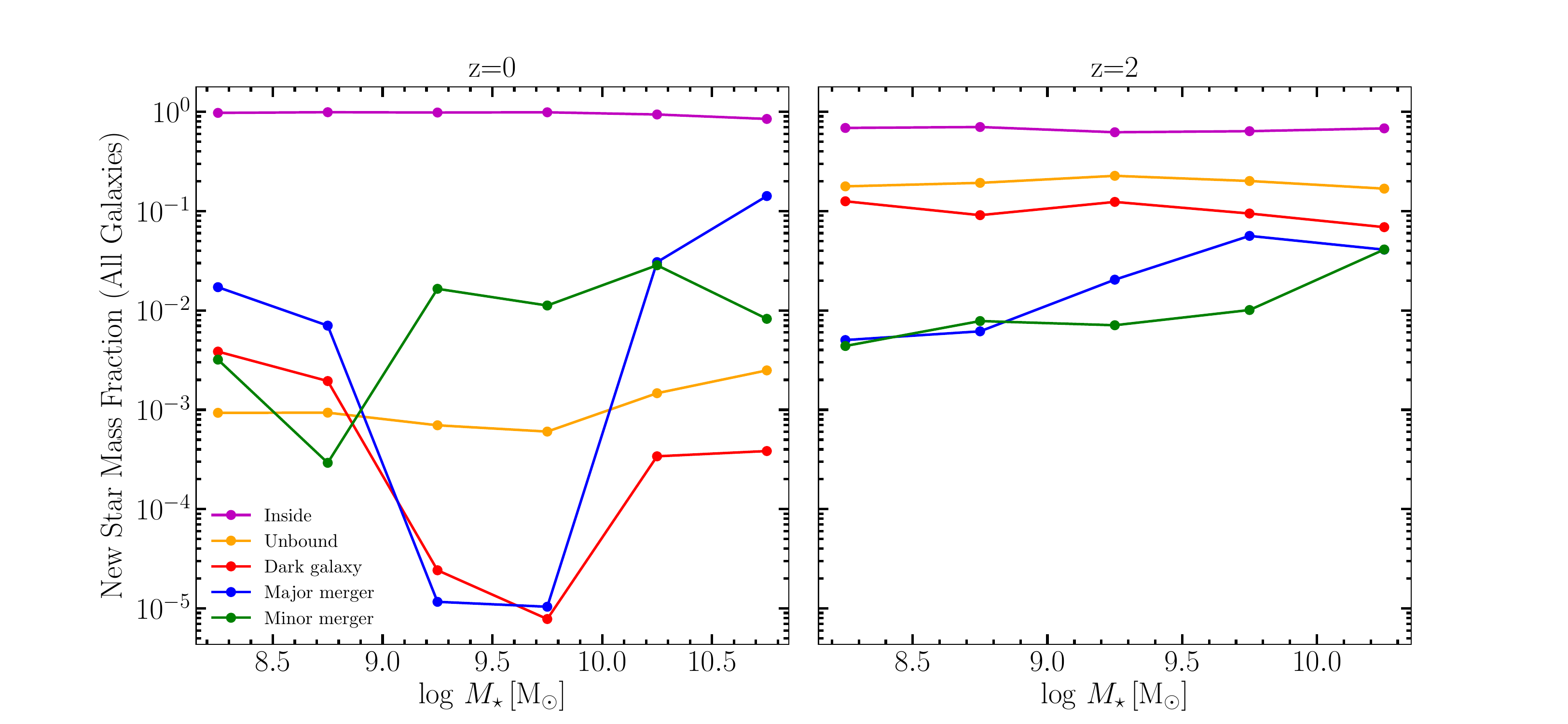}
    \caption{Origin of the gas used to form new stars versus galaxy stellar mass. The galaxies are divided in log-spaced galaxy stellar mass bins of 0.5 dex. For all the galaxies in each bin, the bullets connected by lines indicate the sum of their masses in new stars, normalized by the total mass of newborn stars in the bin (i.e., adding together all categories of all galaxies). The parameter is defined in Eq.~(\ref{eq:jorge1}). Left panel: centrals at $z=0$ with the new stars formed between $z=0.1$ and $z=0$, i.e., stars younger than 1.34 Gyr. Right panel: centrals at $z=2$ with the new stars formed between $z=3$ and $z=2$, i.e., stars younger than 1.13 Gyr. The color code of the lines indicate the location of the gas transformed into new stars: gravitationally bound to the original galaxy (violet line), unbound (orange line), bound to a {\em dark} galaxy (red), or accreted through major (blue) or minor (green) mergers.
    }
    \label{fig:massinit}
  \end{figure*}
\section{Gas fuelling recent star formation at different epochs}
\label{res:1}
As we argued in Section~\ref{intro}, understanding the origin of the star-forming gas is fundamental to  use observational data as diagnostic tools. The example of one such observation, suggestive of metal-poor gas accretion, was examined in Section~\ref{res:2}.  
In this other section, we aim at understanding the origin and metal content of the gas fuelling recent star formation in galaxies.

\subsection{Origin of the gas fuelling star formation}
We analyse the source of gas that recently formed stars at low ($z=0$) and high ($z=2$) redshifts, tracing back in time stellar particles to identify their {\em parent} gas particles (details are given in Section~\ref{subsection:partanalysis}). We employ all central galaxies in the simulation to improve statistics.
This approach generally suffices for our purpose, however, the lack of major mergers and mergers with dark haloes at $z=0$ (left panel in Figure~\ref{fig:num}) results in scarcely populated bins and large bin-to-bin variations. The statistics could be improved by using the largest {\sc eagle} volume at the cost of not using the high spatial and mass resolution required for the analysis in Section~\ref{res:2}, or employing a galaxy sample that does not match that analysis. We therefore preferred to keep the same sample of galaxies\footnote{However, tests carried out using the largest simulation of the {\sc eagle} suite yield results consistent with those described below.}. We warn the reader that, in what follows, the redshift zero statistics of major mergers and mergers with dark haloes has to be taken with caution. We will generally plot the dispersion around the mean value, although bins containing only one or a couple of galaxies should have very large errors assigned. 

For each galaxy of the $z=0$ sample, we identify the location of the gas at $z=0.1$ that corresponds to stars formed between $z=0.1$ and $z=0$, i.e., stars younger than 1.34 Gyr. The newly-formed stars are classified into different categories according to the location of the parent gas at $z=0.1$ (Table~\ref{tab:cat}). Analogously, the exercise was repeated at $z=2$, studying the gas that corresponds to stars formed in between $z=3$ and $z=2$, i.e., stars younger than 1.13 Gyr at that time. As we explain in Section~\ref{subsection:partanalysis}, the redshift intervals were selected to have similar time lapses between snapshots.
We divide the samples in log-spaced galaxy stellar mass bins of 0.5 dex. For each category of parent gas (Table~\ref{tab:cat}), we compute the mass of all star particles it generates. These masses, normalized by the total initial mass in the mass bin, are shown in Figure \ref{fig:massinit}. 
Specifically, if $m_{\rm new\star}^{g,j}$ is the mass in new stars of the $j$-th galaxy coming from gas in the $g$-th category, we plot
\begin{equation}
f_{\rm new\star}^g(M_\star)=\frac{\sum_{\forall j}\,m_{\rm new\star}^{g,j}\,\Pi\Big(\frac{\log M_\star^j-\log M_\star}{0.5}\Big)}{\sum_{\forall g', j}\,m_{\rm new\star}^{g',j}\,\Pi\Big(\frac{\log M_\star^j-\log M_\star}{0.5}\Big)},
\label{eq:jorge1}
\end{equation}
where $M_\star^j$ is the total stellar mass of the $j$-th galaxy, $M_\star$ is the stellar mass of a particular stellar mass bin,  and $\Pi(x)$ is the top-hat function,
\begin{equation}
\Pi(x)=
    \begin{cases}
    1 & -1/2\le x \le 1/2,\\
    0 & {\rm elsewhere}.
   \end{cases}
\end{equation}
The left panel in Figure~\ref{fig:massinit} corresponds to central galaxies at $z=0$ whereas the right panel shows galaxies at $z=2$. The color code of the lines refers to $g$, i.e., the original category of the gas transformed into new stars: gravitationally bound to the original galaxy (violet line), unbound (orange line), bound to a dark galaxy (red line), or accreted through major (blue line) or minor (green line) mergers. 
For galaxies at $z=0$ (Figure~\ref{fig:massinit}, left panel), the contribution from pre-existing gas in the galaxy is significantly higher than from accreted gas (all the other categories together), since the latter corresponds to less than about 2\,\% of the total mass for $M_\star<10^{10.5}\ \mathrm{M_{\odot}}$. The importance of star formation due to major mergers increases in galaxies with $M_\star>10^{10.5}\ \mathrm{M_{\odot}}$ as it contributes to around 10\,\% of the initial mass. The situation changes at high redshift  (Figure~\ref{fig:massinit}, right panel). Although stars formed from pre-existing gas in the galaxy are still dominant, the contribution from accreted gas is considerably higher than at low redshift. More than 10\,\% of the initial mass in new stars corresponds to gas that was not bound to any galaxy, and about 10\,\% comes from mergers with dark galaxies (which primarily contain gas and dark matter). For  high-mass galaxies, the contribution from minor and major mergers is higher (as at $z=0$). In general, the variation with stellar mass is smoother than at low redshift and the contribution from minor and major mergers clearly increases with galaxy mass.
\begin{figure*}
    \centering
    \includegraphics[width=1.\textwidth]{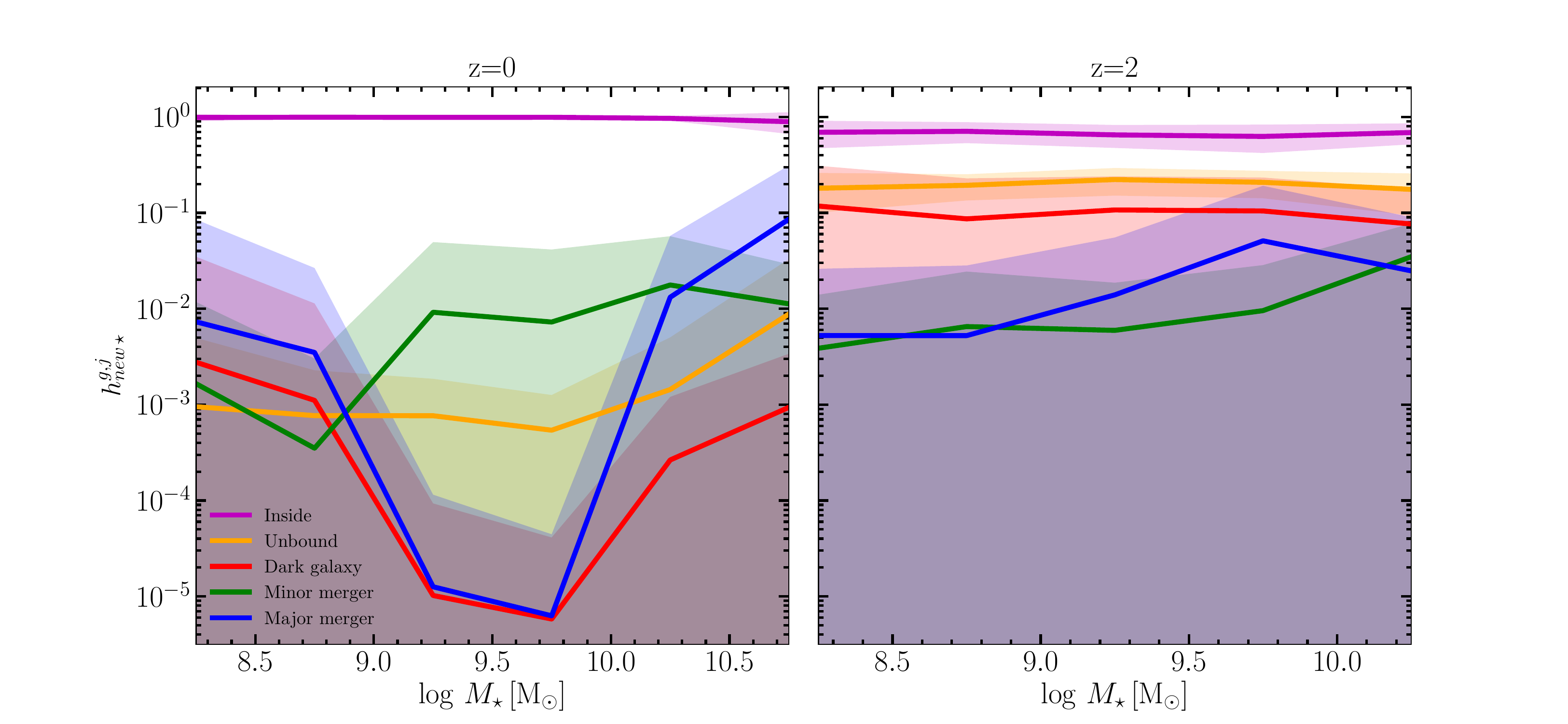}
    \caption{Mass fraction of new stars relative to the total mass of new stars in each individual galaxy. The fraction $h_{\rm new\star}^{g,j}$ is defined in Eq.~(\ref{eq:jorge2}). Each category for the origin of the gas is color coded as in Figure~\ref{fig:massinit}, with the solid lines indicating mean values and the colored regions $\pm$ one standard deviation.  Left panel: centrals at $z=0$ with the new stars formed between $z=0.1$ and $z=0$. Right panel: centrals at $z=2$ with the new stars formed between $z=3$ and $z=2$. 
    }
    \label{fig:massnew_var}
\end{figure*}
\begin{figure*}
    \centering
   \includegraphics[width=1.\textwidth]{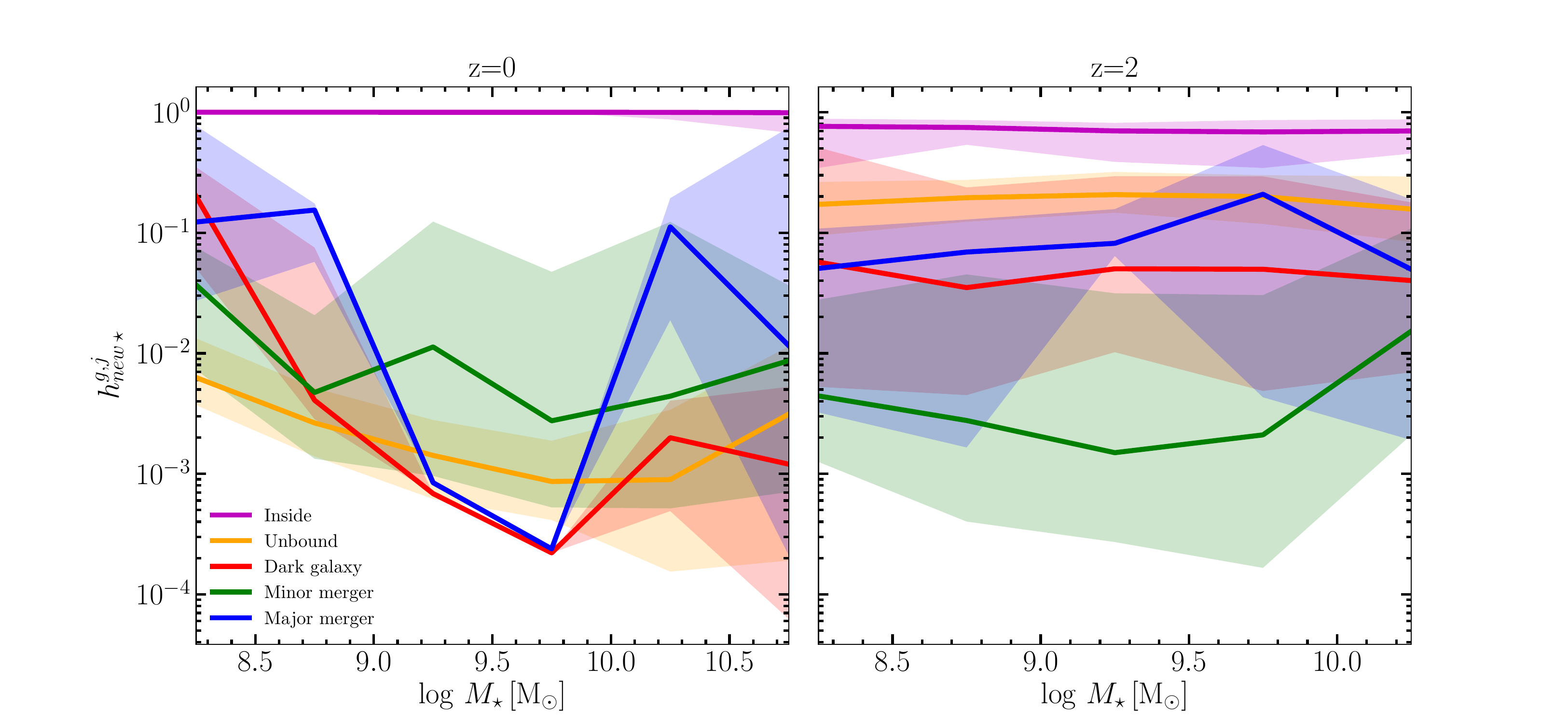}
    \caption{Mass fraction of new stars in a category relative to the total mass of new stars. The figure is similar to Figure~\ref{fig:massnew_var}, except that galaxies with no contribution of gas with a given category are excluded from the statistics, and mean and standard deviations have been replaced with median and $\mathrm{10^{th}}$ and $\mathrm{90^{th}}$ percentiles, respectively. Refer to the caption of Figure~\ref{fig:massnew_var} for further details. 
    }
    \label{fig:massnew}
  \end{figure*}
The fraction represented in Figure~\ref{fig:massinit} (Eq.~[\ref{eq:jorge1}]) quantifies the properties of all galaxies taken together, however, it does not inform us about the behaviour of individual galaxies. In other words, it does not tell us whether all galaxies behave the same or whether a few galaxies have had a large contribution from external gas. In order to distinguish between the two possibilities, the mass fraction for individual galaxies, 
\begin{equation}
h_{\rm new\star}^{g,j}=\frac{m_{\rm new\star}^{g,j}}{\sum_{\forall g'}\,m_{\rm new\star}^{g',j}},
\label{eq:jorge2}
\end{equation}
was also computed to study its statistical properties.
Figure \ref{fig:massnew_var} shows the mean of these mass fractions for each category, considering all the galaxies in each mass bin (including the ones that have $m_{\rm new\star}^{g,j}=0$ in that category). Left and right panels correspond to the analyses at redshifts $z=0$ and $z=2$, respectively, and the color code is the same as the one used in Figure~\ref{fig:massinit}. For each category, the solid lines correspond to the mean of the distribution and the colored regions indicate the standard deviation. Note that for the bins with only a few galaxies (Figure~\ref{fig:num}), this standard deviation is almost certainly underestimating the true scatter since it does not include any estimate of the foreseeable counting error. As expected, the means show a trend similar to the one seen in Figure~\ref{fig:massinit}, however, the scatter is significant. In the left panel of Figure \ref{fig:massnew_var}, we see that, at low redshift, the major contribution to the mass in new stars is due to pre-existing gas in the galaxy.  Nonetheless, within one standard deviation of the average properties, some galaxies have external gas contributions as large as 30\,\% (major mergers at the high-mass end; Figure~\ref{fig:massnew_var}, left panel), and the typical external contribution is as large as 10\,\%. In the case of intermediate mass galaxies, this contribution comes from minor mergers. Assuming the mass ratio to follow a Gaussian distribution, around 16\,\% of the galaxies have external gas contributions larger than 10\,\%. In low mass galaxies, the external gas comes from major mergers and mergers with dark galaxies.
The right panel in Figure~\ref{fig:massnew_var} shows the mass fractions at redshift $\simeq 2$. Although pre-existing gas still dominates the recent star formation, unbound gas and gas coming from mergers with dark galaxies have a considerably higher contribution than at $z=0$ in the whole mass range. Often the external gas contribution (i.e., all origins but gravitationally bound gas) reaches 30\,\%.

The statistical properties displayed in Figure~\ref{fig:massnew_var} are strongly biased towards galaxies having no stars formed from a particular category of gas, i.e., those for which $m_{\rm new\star}^{g,j}=0$ for a particular $g$. We remove them when computing Figure~\ref{fig:massnew}. In addition, to become independent of the shape of the distribution function of the fraction, the figure represents medians and percentiles rather than the means and standard deviations used in Figure~\ref{fig:massnew_var}. For each category, the solid line indicates the median and the colored region embraces the $\mathrm{10^{th}}$ and $\mathrm{90^{th}}$ percentiles.  
As in the previous figures, left and right panels correspond to redshifts $z=0$ and $z=2$, respectively, with the color code being the same as the one used in Figures \ref{fig:massinit} and \ref{fig:massnew_var}.  Despite the fact that at $z=0$ the main contribution corresponds to gas that was already in the galaxy, for the galaxies that have received gas from mergers with dark galaxies (red line and red area) and major mergers (blue line and blue area), the external gas makes about 10\,\% of the new star mass at the low-mass end. Taking into account the scatter, for some 10\,\% of the galaxies, the contribution from external gas becomes 60\,\% if the gas comes from major megers,  and 20\,\% if the gas was obtained from mergers with dark galaxies. At high redshift the trends are the same but amplified. Unbound gas contributes with more than the 10\,\% of the mass of new stars. In some systems, the fraction of new stars formed from gas coming from mergers with dark galaxies becomes 40\,\%.  

The question arises as to whether the mass in newly formed stars makes up a significant contribution to the total stellar mass of the galaxies. Considering all categories together at  $z=0$, the new stars contribute with 20\,\% of the total mass at low mass and 10\,\% at high mass. The fraction increases to around 50\,\% at redshift 2. As for the trends and relative contribution of the various categories, they are almost identical to that shown in Figure~\ref{fig:massnew} and discussed in the previous paragraph.

\begin{figure*}
    \centering
     \includegraphics[width=1.\textwidth]{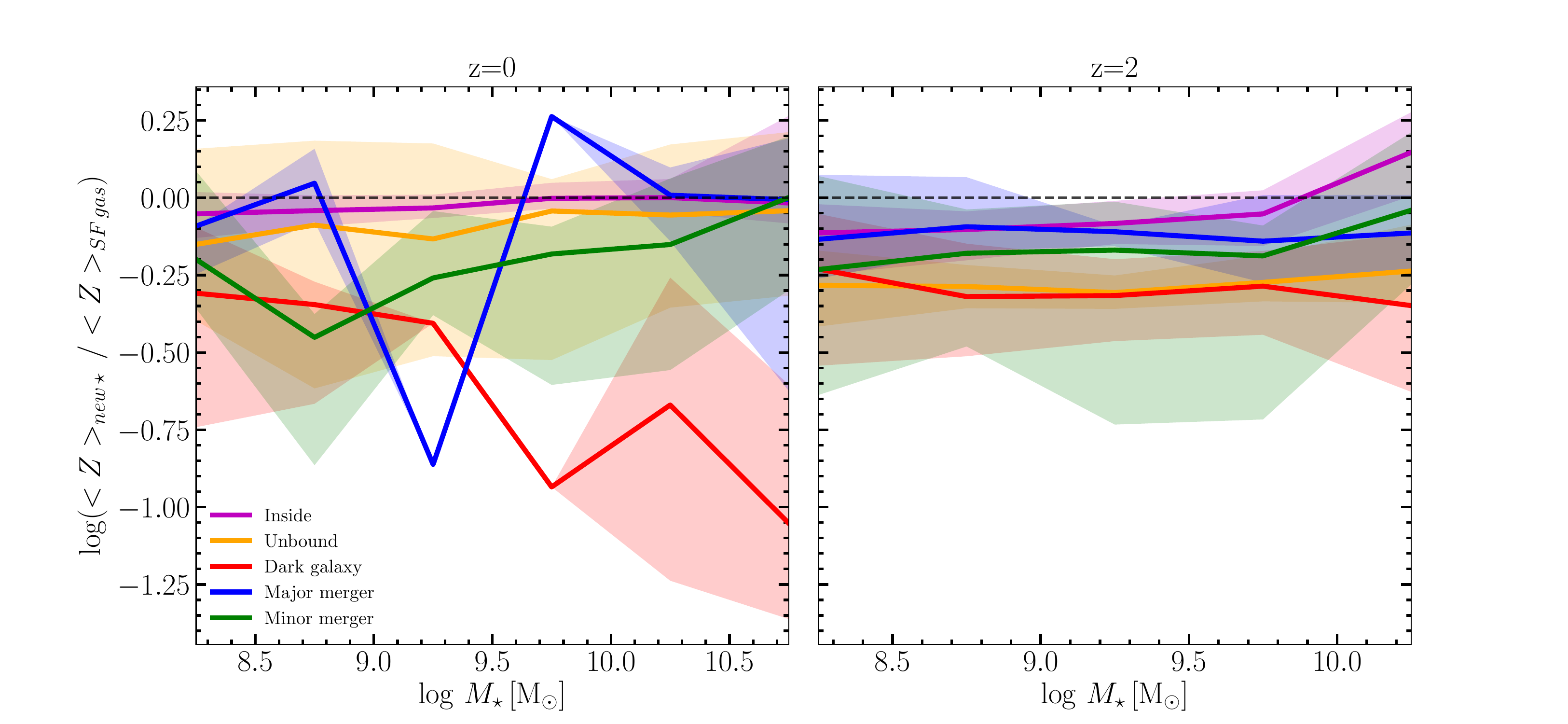}
    \caption{Mean metallicity of the new stars relative to the mean metallicity of the star-forming gas of the galaxy at present. Only galaxies with newly formed stars are considered. The color code and the labels are the same as in Figure~\ref{fig:massnew_var}, and we refer to its caption for details.  The black dashed lines show  metallicity ratio equal to 1 and was included for reference.
    }
    \label{fig:met}
\end{figure*}
\subsection{Metallicity of the new stars}
\label{sec:metallicity}
We also examine how the metallicity of the new stars varies depending on the source of the gas they formed from. For each galaxy, we calculate the mean metallicity of the new stars which is then referred to the mean metallicity of the star-forming gas of the galaxy at present. Figure \ref{fig:met} shows the median of this metallicity ratio for each category and stellar mass bins.  The layout and color code are the same as in Figures \ref{fig:massinit} and \ref{fig:massnew}. At $z=0$, the metallicity of the stars formed from pre-existing gas is quite similar to the metallicity of the star-forming gas in the whole mass range, and with a scatter considerably lower than the metallicities from gas accreted to the galaxy. In contrast, the metallicity of the stars formed from gas that was previously bound to dark galaxies is considerably lower than the mean metallicity of the star-forming gas. We also observe that it decreases with increasing galaxy mass for reasons which are not clear at present. The metallicity of the contribution of minor mergers is also smaller, although it increases at the low and high-mass ends. The metallicity of major mergers is quite similar to the metallicity of the new stars formed from pre-existing gas. Intermediate-mass galaxies have a behaviour quite irregular, which we attribute to the fact that only few galaxies contribute to this category in this stellar mass range (see Figure~\ref{fig:num}), thus leading to a large Poisson noise. At $z=0$, the metallicities of the stars formed from unbound gas is quite high and with very large scatter, with a median only slightly lower than the one corresponding to gas already in the galaxy. On the other hand, at $z=2$, the metallicites of the stars formed from pre-existing gas and from gas coming from major mergers are quite similar, both slightly lower than the metallicity of the star-forming gas in the galaxies. The gas from  major mergers shows higher metallicities for high-mass galaxies, which might indicate that these stars are formed from recycled gas. Despite the metallicity of minor mergers is slightly lower, it also increases for high-mass galaxies. The metallicites of the stars that are formed out of unbound gas or gas coming from mergers with gas-rich dark galaxies are slightly lower than the one of the stars formed from pre-existing gas. 
We also see that the variation with stellar mass is smoother at high redshift.

Two additional conclusions can be drawn from the above results. In low-mass galaxies, recent star formation due to gas coming from mergers with gas-rich dark galaxies has a metallicity distribution that is very broad, indicating that there are galaxies whose newly formed stars have considerably lower metallicities than the metallicity of the star-forming gas. 
Second,  the stars formed from unbound gas at  $z=0$ have high metallicity, which indicates that this gas comes from galactic fountains, where material pre-processed in stars is ejected to the CGM or even the IGM, and then cools down to be re-accreted onto the galaxy disc \citep[e.g.,][]{2017ASSL..430..323F}. Thus, fast galactic winds are operationally classified as gravitationally unbound to the galaxy even though they may be part of the CGM.


\section{Conclusions}
\label{sect:conclusions}

We have studied the origin of the gas forming new stars in galaxies produced by the state-of-the-art cosmological hydrodynamical simulation \textsc{eagle}. The processes of external gas accretion is thought to be fundamental but only loosely constrained observationally, and our effort aims at improving the diagnostic capabilities when interpreting observations.

The process of acquiring  external gas to form new stars is expected to be localized and intermittent (Section~\ref{intro}). Thus, together with inefficiently mixing with pre-existing gas, it should leave imprint in the spatial distribution of gas and metallicity within galaxies. We synthesize spatially-resolved star-formation distributions to compare them with observations of star-forming disc galaxies. Specifically, we analyzed the local relation between gas-phase metallicity and star formation rate in the subsample of 107 central galaxies at $z=0$ that have indisputable {\em disc-like} features in \textsc{eagle}. We generate $1~\mathrm{kpc^2}$ resolution maps of $Z_g$ and $\Sigma_{\rm SFR}$ for the galaxies projected in face-on configuration.
The maps thus constructed show regions in the outskirts of massive galaxies ($\gtrsim   10^{10}~{\rm M}_\odot$) where the larger the $\Sigma_\mathrm{SFR}$ the smaller $Z_g$, with the relation reversing sign in the central parts of the galaxies (Figure \ref{fig:g2}). For lower mass galaxies, the anti-correlation between $\mathrm{\Sigma_{SFR}}$ and $Z_g$  is even stronger and remains negative all the way from the ourkirsts to the center (Figure \ref{fig:g4}). 
%
We employ these maps to quantify the local relation between gas-phase metallicity and star formation rate already observed in discs of nearby star-forming galaxies (Section~\ref{intro}). Separated by stellar mass, we evaluate the slope of the scatter plot  $Z_g$ versus $\mathrm{\Sigma_{SFR}}$, which was then compared with the observed  relation found by \citet{2019ApJ...882....9S} studying 736 nearby spiral galaxies from the MaNGA survey.
The trend with stellar mass of the \textsc{eagle} galaxies is close to the one observed in MaNGA. Low-mass galaxies have negative slopes and hence, an anti-correlation between $Z_g$ and $\mathrm{\Sigma_{SFR}}$ whereas, at the high-mass end, the slope reverses sign denoting positive correlation (Figure \ref{fig:slopes}).  The local $Z_g$ --$\mathrm{\Sigma_{SFR}}$ anti-correlation present in {\sc eagle} galaxies is consistent with external metal-poor gas triggering star formation locally. This scenario was already proposed in literature to explain the presence of regions with high SFR and low metallicity in nearby low-mass galaxies \citep{2015ApJ...810L..15S,2018MNRAS.476.4765S,2019ApJ...872..144H}. In the case of higher mass galaxies, where the correlation is positive, \citet{2019ApJ...882....9S} argue that metal-enriched gas, pre-processed in stars, drives star-formation. A very strong dependence of the star-formation efficiency on metallicity could also explain a positive correlation. 

We then study the origin of the gas forming new stars in the  \textsc{eagle} galaxies.  We analyze the source of gas using all galaxies with enough particles in the highest resolution simulation among those of the \textsc{eagle} suite. The analysis was carried out at two particularly revealing epochs: now ($z=0$) and during the peak star formation in the early Univese ($z=2$). Thus, we selected 417 central galaxies at $z=0$, with $10^{8} < M_{\star}/\mathrm{M_{\odot}} \leq 10^{11}$, and 338 centrals at $z=2$, with $10^{8} < M_{\star}/\mathrm{M_{\odot}} \leq 10^{10.5}$. We identified as new stars those formed between $z_{\rm prev}$ and $z$, where $z_{\rm prev}$ was chosen as the snapshot closest to $z=0$ available in the {\sc eagle} database, which corresponds $z=0.1$ and which selects stars younger than 1.3 Gyr.
$z_{\rm prev}=3$ at $z=2$, chosen to grant that the time interval between snapshots is similar at low and high redshift (1.1 Gyr in this case).
According to the location of the gas at $z_{\rm prev}$, we distinguish five different categories: (1) gas gravitationally bound to the galaxy, (2) gas not bound to any galaxy, (3) gas from a different galaxy that merged as a minor merger, (4) gas from a different galaxy that merged as a major merger and, finally, (5) gas bound to a DM halo without stars (dubbed {\em dark galaxy}). Table~\ref{tab:cat} provides a summary with the properties and definition of the different categories. Our results can be condensed as follows:
\begin{itemize}
\item [-] For galaxies of all masses at $z=0$, the contribution to new stars from gas that was already in the galaxy is substantially higher than the one from stars formed from accreted gas, although the importance of major mergers increases at the high-mass end (left panel of Figure \ref{fig:massinit}). In contrast, although stars formed from pre-existing gas are still dominant at $z=2$, accreted gas contributes significantly more than at later times, with more than 10\,\% of the new stars forming from gas that was not bound to any galaxy, and with  roughly 10\,\% forming from gas accreted through mergers with {\em dark} galaxies. Minor and major mergers become more important with increasing galaxy mass at $z=2$. The dependence of the mass fraction with stellar mass is smoother at $z=2$ (cf. left and right panels in Figure \ref{fig:massinit}).
\item [-]
  The scatter in the source of gas among individual galaxies of the same stellar mass is significant. Given the dispersion around the median of the distribution (left panel of Figure \ref{fig:massnew}), even at $z=0$  some low-mass objects have  a large fraction of their newly formed stars produced from accreted gas.  The large scatter persists $z=2$,  and so does the chances for some galaxies to have most of their newly formed mass produced from external gas.

\item [-]
  At $z=0$, the stars formed from pre-existing gas and the gas forming stars at present have quite similar metallicites over the whole mass range. Conversely, the metallicity of the stars formed from gas accreted in mergers with dark galaxies is considerably lower and seems to anti-correlate with the galaxy mass. Minor mergers also contribute with lower metallicities, although the metallicity of this accreted gas is relatively higher at the low and high-mass ends. New stars in low-mass and high-mass galaxies formed from major mergers and pre-existing gas have similar metallicites.
  Trends are similar but smoother at $z=2$ (Figure~\ref{fig:met}).

\item [-]
  A fraction of the gas classified as unbound gas at $z=0$ may come from galactic fountains, as the stars formed from this gas have noticeably high metallicities (Figure~\ref{fig:met}). This is not the case at $z=2$ where the metallicities are low as expected for IGM gas. 

\item[-]
The spatially resolved properties shown in Figures~\ref{fig:g2}\,--\,\ref{fig:g4} can be understood in terms of the integrated galaxy properties. Most new stars are produced out of gas that was already in the halo or from mergers, both in the low-mass and the high-mass ends of the galaxy mass distribution (see Figure~\ref{fig:massnew}, left panel). The metallicity of the stars produced from this gas is slighly higher than average in the high-mass end and clearly lower than average in the low-mass end (see Figure~\ref{fig:met}). Thus, any excess in $\Sigma_{\rm SFR}$ tends to be metal rich in high-mass galaxies and metal-poor in low mass galaxies, which is the trend detected in spatially integrated galaxies (Figure~\ref{fig:slopes}). 
\end{itemize}



\section*{Acknowledgements}

We thank the Virgo Consortium for making their simulation data available. The {\sc eagle} simulations were performed using the DiRAC-2 facility at Durham, managed by the ICC, and the PRACE facility Curie based in France at TGCC, CEA, Bruyeresle-Ch\^atel.
Thanks are due to an anonymous referee for helping us to clarify some of the passages of the paper.
LSD acknowledges support from grant PID2019-107427GB-C32 from The Spanish Ministry of Science and Innovation.
JSA acknowledges support from the Spanish Ministry of Science and Innovation, project  PID2019-107408GB-C43 (ESTALLIDOS), and from Gobierno de Canarias through EU FEDER funding, project PID2020010050.
CDV acknowledges support through grants RYC-2015-18078 and PGC2018-094975-C22 from the Spanish Ministry of  Science and Innovation.

\section*{Data Availability}


All \textsc{eagle} snapshots used in this work are  publicly available  at \url{http://icc.dur.ac.uk/Eagle} . See \citet{2017arXiv170609899T}  for how to access particle data and \citet{2016A&C....15...72M} for information about the EAGLE database.


\newcommand\ac{Astron.~Comp.}
\newcommand\aar{A\& ARev}
\newcommand\noopsort[1]{#1}
\bibliographystyle{mnras}








\bsp	
\label{lastpage}
\end{document}